\documentclass[12pt]{article}
\usepackage[dvips]{graphics}
\usepackage{epsfig}
\usepackage{epsf}

\usepackage{scicite}
\usepackage{times}

\newenvironment{sciabstract}{%
\begin{quote} \bf}
{\end{quote}}

\newcounter{lastnote}
\newenvironment{scilastnote}{%
\setcounter{lastnote}{\value{enumiv}}%
\addtocounter{lastnote}{+1}%
\begin{list}%
{\arabic{lastnote}.}
{\setlength{\leftmargin}{.22in}}
{\setlength{\labelsep}{.5em}}}
{\end{list}}

\newcommand{\oversim}[2]{\protect{\mbox{\lower0.5ex\vbox{%
   \baselineskip=0pt\lineskip=0.2ex
   \ialign{$\mathsurround=0pt #1\hfil##\hfil$\crcr#2\crcr\sim\crcr}}}}} 
\newcommand{\simgreat}{\mbox{$\,\mathrel{\mathpalette\oversim>}\,$}} 
\newcommand{\simless} {\mbox{$\,\mathrel{\mathpalette\oversim<}\,$}} 

\title{The Initial Mass Function of Stars:\\
Evidence for Uniformity in Variable Systems}

\author{
\normalsize{ --- {\cal SCIENCE}, 4th January 2002, Vol.~295, No.~5552,
p.~82--91 --- }\\
\footnotesize{http://www.sciencemag.org/content/}
\vspace{-3mm}\\
\footnotesize{The published paper has electronic on-line-tables only.}
\vspace{10mm}\\
Pavel Kroupa
\vspace{0mm}\\
\normalsize{Institut f\"ur Theoretische Physik und Astrophysik,
Universit\"at Kiel}\\
\normalsize{D-24098 Kiel, Germany}\\
\\
\normalsize{E-mail: pavel@astrophysik.uni-kiel.de}
}

\date{}

\begin{document} 


\baselineskip12pt


\maketitle 


\begin{sciabstract}
The distribution of stellar masses that form in one star-formation
event in a given volume of space is called the initial mass function
(IMF). The IMF has been estimated from low-mass brown dwarfs to very
massive stars. Combining IMF estimates for different populations in
which the stars can be observed individually unveils an extraordinary
uniformity of the IMF. This general insight appears to hold for
populations including present-day star formation in small molecular
clouds, rich and dense massive star-clusters forming in giant clouds,
through to ancient and metal-poor exotic stellar populations that may
be dominated by dark matter.  This apparent universality of the IMF is
a challenge for star formation theory because elementary
considerations suggest that the IMF ought to systematically vary with
star-forming conditions.
\end{sciabstract}



The physics of star formation determines the conversion of gas to
stars.  The outcome of star formation are stars with a range of
masses. Astrophysicists refer to the distribution of stellar masses as
the stellar initial mass function.  Together with the time-modulation
of the star-formation rate, the IMF dictates the evolution and fate of
galaxies and star clusters.  The evolution of a stellar system is
driven by the relative initial numbers of brown dwarfs (BDs, $\simless
0.072\,M_\odot$) that do not fuse H to He, very low-mass stars ($0.072
- 0.5\,M_\odot$), low-mass stars ($0.5 - 1\,M_\odot$),
intermediate-mass stars ($1-8\,M_\odot$) and massive stars
($m>8\,M_\odot$). Non-luminous BDs through to dim low-mass stars
remove gas from the interstellar medium (ISM), locking-up an
increasing amount of the mass of galaxies over cosmological time
scales. Intermediate and luminous but short-lived massive stars expel
a large fraction of their mass when they die and thereby enrich the
ISM with elements heavier than H and He. They heat the ISM through
radiation, outflows, winds and supernovae \cite{CSc,Hens}. It is
therefore of much importance to quantify the relative numbers of stars
in different mass ranges and to find systematic variations of the IMF
with different star-forming conditions.  Identifying systematic
variations of star formation would allow us to understand the physics
involved in assembling each of the mass ranges, and thus to probe
early cosmological events.  Determining the IMF of a stellar
population with mixed ages is a difficult problem. Stellar masses
cannot be weighed directly in most instances \cite{kepler} so the mass
has to be deduced indirectly by measuring the star's luminosity and
evolutionary state.

The history of the subject began in 1955 at the Australian National
University when Edwin~E. Salpeter published the first estimate
\cite{S55} of the IMF for stars in the solar-neighborhood
\cite{neighb}.  For stars with masses in the range $0.4-10\,M_\odot$
he found that it can be described by a power-law form with an index
$\alpha=2.35$.  This result implied a diverging mass density for
$m\rightarrow 0$, which was interesting because dark matter was
speculated, until the early 1990's, to possibly be made-up of faint
stars or sub-stellar objects.  Studies of the stellar velocities in
the solar-neighborhood also implied a large amount of missing, or
dark, mass in the disk of the Milky Way (MW) \cite{B84}.  Beginning in
the early 1950's Wilhelm Gliese in Heidelberg began a careful compilation
of all known stars within the solar neighborhood with accurately known
distance. The edition published in 1969 became known as the famous
{\it Gliese Catalogue of Nearby stars}, the modern version of which
\cite{neighb,JW97} constitutes the most complete and best-studied
stellar sample in existence.  During the early 1980's newly developed
automatic plate-measuring machines made it possible to discriminate
between many distant galaxies and a few nearby main-sequence stars in
the hundred thousand images on a single photographic plate.  This
allowed Neill~Reid and Gerard Gilmore at Edinburgh Observatory to make
photographic surveys of the sky with the aim of finding very low-mass
stars beyond the solar neighborhood \cite{RG82}. Together with the
Gliese Catalogue this survey and others that followed using the same
technique significantly improved knowledge of the space density of
very low-mass stars \cite{MS79,Sc86}.  The form of the IMF for
low-mass stars was further revised in the early 1990's in Cambridge
(UK) through improved theoretical understanding of the
mass--luminosity relation of low mass stars and the evaluation of the
observational errors due to unresolved binary systems
\cite{KTG90,KTG93}, finding confirmation by subsequent work
\cite{GBF}. For massive stars John Scalo's \cite{Sc86} determination
($\alpha\approx 2.7$) in Austin in 1986 remained in use.  It is even
today the most thorough analysis of the IMF in existens.  It is
superseded now by Phillip Massey's \cite{M98} work at Tucson who
demonstrated through extensive spectroscopic classification that
Salpeter's original result extends up to the most massive stars known
to exist with $m\approx120\,M_\odot$.

Today we know that the IMF for solar-neighborhood stars flattens
significantly below about $0.5\,M_\odot$. The IMF for BDs is even
shallower, as shown by Gilles Chabrier at Berkeley in 2001
\cite{Ch01b}, so that very-low mass stars and BDs contribute an
insignificant amount to the local mass density. The need for dark
matter in the MW disk also disappeared as improved kinematical data of
stars in the MW disk became available \cite{Kuijken,FF94}.  Popular
analytical descriptions of the IMF and some definitions are summarized
in Table~\ref{tab:imfs}.

\section*{The Form of the IMF}

Assuming all binary and higher-order stellar systems can be resolved
into individual stars in some population such as the solar
neighborhood \cite{neighb} and that only main-sequence stars are
selected for, then the number of stars per pc$^3$ in the mass interval
$m$ to $m+dm$ is $dN=\Xi(m)\,dm$, where $\Xi(m)$ is the observed
present-day mass function (PDMF). The number of stars per pc$^3$ in
the absolute magnitude \cite{magn} interval $M_P$ to $M_P+dM_P$ is
$dN=-\Psi(M_P)\,dM_P$, where $\Psi(M_P)$ is the stellar luminosity
function (LF). It is constructed by counting the number of stars in
the survey volume per magnitude interval, and $P$ signifies an
observational photometric pass-band such as the $V$-band.  Thus
\begin{equation}
\Xi(m) = -\Psi(M_P)\,(dm/dM_P)^{-1}.
\label{eq:mf_lf}
\end{equation}
Because the derivative of the stellar mass--luminosity relation (MLR),
$m(M_P)=m(M_P,Z,\tau,\mathbf{s})$, is needed to calculate $\Xi(m)$,
any uncertainties in stellar structure and evolution theory on the one
hand or in observational ML-data on the other hand will be
magnified. The dependence of the MLR on the star's chemical
composition, $Z$, its age, $\tau$, and its spin vector $\mathbf{s}$,
is explicitly stated here. This is because stars with fewer metals
(lower opacity) than the Sun are brighter. Main-sequence stars
brighten with time and they lose mass. Rotating stars are dimmer
because of the reduced internal pressure. Mass loss and rotation also
alter the MLR for intermediate and especially high-mass stars
\cite{MPV01}.

The IMF follows by correcting the observed number of main sequence
stars for the number of stars that have evolved off the main sequence.
Defining $t=0$ to be the time when the Galaxy that now has an age
$t=\tau_{\rm G}$ formed, the number of stars per pc$^3$ in the mass
interval $m,m+dm$ that form in the time interval $t,t+dt$ is
$dN=\xi(m,t)\,dm\times b(t)\,dt$. The expected time-dependence of the
IMF is explicitly stated, and $b(t)$ is the time-modulation of the
IMF. This is the normalized star-formation history (SFH), with $(1/\tau_{\rm
G}) \int_0^{\tau_{\rm G}}b(t)\,dt = 1$.  Stars that have main-sequence
life-times $\tau(m) < \tau_{\rm G}$ leave the stellar population
unless they were born during the most recent time interval
$\tau(m)$. The number density of such stars with masses in the range
$m,m+dm$ still on the main sequence and the total number density of
stars with $\tau(m) \ge \tau_{\rm G}$, are, respectively,
\begin{equation}
\Xi(m) = 
   \xi(m){1\over \tau_{\rm G}}
   \left\{ 
   \begin{array}{l@{\quad\quad,\quad}l}
   \int_{\tau_{\rm G}-\tau(m)}^{\tau_{\rm G}} b(t)dt &
   \tau(m) < \tau_{\rm G},\\
   \int_0^{\tau_{\rm G}} b(t)\,dt & \tau(m) \ge \tau_{\rm G},
   \end{array}\right.
\label{eq:imf_pdmf}
\end{equation}
where the time-averaged IMF, $\xi(m)$, has now been defined. Thus, for
low-mass stars $\Xi=\xi$, while for a sub-population of massive stars
that has an age $\Delta t \ll \tau_{\rm G}$, $\Xi=\xi\,(\Delta
t/\tau_{\rm G})$ for those stars of mass $m$ for which $\tau(m)>\Delta
t$. This indicates how an observed high-mass IMF in an OB association,
for example, is scaled to the Galactic-field \cite{field} IMF for
low-mass stars.  In this case the different spatial distribution by
different disk-scale heights of old and young stars also needs to be
taken into account, which is done globally by calculating the stellar
surface density in the MW disk \cite{MS79,Sc86}.  In a star cluster or
association with an age $\tau_{\rm cl}\ll\tau_{\rm G}$, $\tau_{\rm
cl}$ replaces $\tau_{\rm G}$ in eq.~\ref{eq:imf_pdmf}.  Examples of
the time-modulation of the IMF are $b(t)=1$ (constant star-formation
rate) or a Dirac-delta function, $b(t)=\tau_{\rm cl} \times
\delta(t-t_0)$ (all stars formed at the same time $t_0$).

\paragraph*{Massive stars}

Studying the distribution of massive stars is complicated because most
of their energy is emitted at far-UV wavelengths that are not
accessible from Earth, and they have short main-sequence life-times
\cite{M98}.  For example, an $85\,M_\odot$ star cannot be
distinguished from a $40\,M_\odot$ star on the basis of $M_V$ alone.
Constructing $\Psi(M_V)$ to get $\Xi(m)$ for a mixed-age population
does not work if optical or even UV-bands are used.  Instead, spectral
classification and broad-band photometry for estimation of the
reddening of the star-light through interstellar dust has to be
performed on a star-by-star basis to measure the effective
temperature, $T_{\rm eff}$, and the bolometric magnitude, $M_{\rm
bol}$, from which $m$ is obtained, allowing the construction of
$\Xi(m)$.

Having obtained $\Xi(m)$ for a population, the IMF follows by applying
eq.~\ref{eq:imf_pdmf}. Studies that rely on broad-band optical
photometry consistently arrive at IMFs that are steeper with a
power-law index $\alpha_3\approx3$ (see eq.~\ref{eq:imf_mult} below),
rather than $\alpha_3=2.2\pm0.1$ consistently found using spectral
classification for a wide range of stellar populations
\cite{M98}. However, multiple systems that are not resolved into
individual stellar companions hide their less-luminous members. This
is a serious problem because observations have shown that most massive
stars are in binary and higher-order multiple systems
\cite{Duchene01,Preibisch99}. Correcting for the missed companions
leads to systematically larger $\alpha_3\approx2.7$ values
\cite{SR91}.  The larger value, $\alpha \approx 3\pm0.1$, is also
suggested by a completely independent but indirect approach relying on
the distribution of ultra-compact HII regions in the MW \cite{Cass00}.

Massive main-sequence stars have substantial winds flowing outwards
with velocities of a few~100 to a few~1000~km/s \cite{KP00}, but they
do not loose more than about 10~per cent of their mass
\cite{Garcia96a,Garcia96b}. More problematic is that these massive
stars are rapidly rotating when they form and so are sub-luminous as a
result of reduced internal pressure. They decelerate 
during their main-sequence life-time owing to the angular-momentum
loss through their winds and become more luminous more rapidly than
non-rotating stars \cite{MM00}.  The mass--luminosity relation for a
population of stars that have a range of ages is therefore broadened
making mass estimates from $M_{\rm bol}$ uncertain by up to 50~\%
\cite{MPV01}, a source of error also not yet taken into account in the
derivations of the IMF. Another problem is that
$m\simgreat40\,M_\odot$ stars may finish their assembly after burning
a significant proportion of their central~H so that a zero-age-main
sequence may not exist for massive stars \cite{MB01}.

\paragraph*{Intermediate-mass stars}

These stars have main-sequence life-times similar to the age of the MW
disk. Solving equation~\ref{eq:imf_pdmf} becomes sensitive to the SFH
of the solar neighborhood and to the age and structure of the
disk. None of these are known very well. Conversion of the PDMF to the
IMF also depends on corrections for evolution along the main sequence
if the ages of the stars were known.  Deriving the IMF for
intermediate-mass solar-neighborhood stars is therefore subject to
difficulties that do not allow an unambiguous estimate of the IMF
\cite{Binney00}.  The gap between massive and low-mass stars is
bridged by assuming the IMF is continuous and differentiable.

\paragraph*{Low-mass and very-low-mass stars in the Galactic field}

Galactic-field stars \cite{field} have an average age of about 5~Ga
and represent a mixture of many star-formation events. The IMF deduced
for these is therefore a time-averaged IMF which is an interesting
quantity for at least two reasons, namely for the mass-budget of the
MW disk, and as a bench-mark against which the IMFs measured in
presently occurring star-formation events can be compared with to
distill possible variations about the mean.

There are two well-tried approaches to determine $\Psi(M_V)$ in
eq.~\ref{eq:mf_lf} for Galactic-field stars.  The first and most
straightforward method for estimating the IMF consists of creating a
local volume-limited catalogue of nearby stars with accurate distance
measurements.  The second method is to make deep pencil-beam surveys
to extract a few hundred low-mass stars from a hundred-thousand
stellar and galactic images.  This approach leads to larger stellar
samples because many lines-of-sight into the Galactic field ranging to
distances of a few~100~pc to a~few~kpc are possible \cite{LFphot}.
The local {\it nearby LF}, $\Psi_{\rm near}$, and the deep {\it
photometric LF}, $\Psi_{\rm phot}$, are displayed in
Fig.~\ref{fig:lfs}. They differ significantly for stars fainter than
$M_V\approx11.5$ causing controversy in the past \cite{LFdiscr}.  The
solar neighborhood sample cannot have a spurious but statistically
significant over-abundance of very-low-mass stars because the velocity
dispersion in the disk is large, $\approx30$~pc/My. Any significant
overabundance of stars within a sphere with a radius of 30~pc would
disappear within one~My, and cannot be created nor sustained by any
physically plausible mechanism in a population of stars with stellar
ages spanning the age of the MW disk.

The slope of the MLR (Fig.~\ref{fig:mlr}) is very small at faint
luminosities leading to large uncertainties in the MF near the
hydrogen burning mass limit ($\approx 0.072\,M_\odot$, \cite{ChBa}).
Any non-linear structure in the MLR is mapped into observable
structure in the LF (eq.~\ref{eq:mf_lf}), provided the MF does not
have compensating structure.  The derivative has a sharp maximum at
$M_V\approx11.5$, this being the origin of the maximum in $\Psi_{\rm
phot}$ near $M_V=12$ \cite{KT97}.

In addition to the non-linearities in the MLR relation unresolved
multiple systems affect the MF derived from $\Psi_{\rm phot}$.  This
is a serious issue because no stellar population is known to exist
that has a binary proportion smaller than 50~\%.  Suppose an observer
sees 100~systems. Of these~40, 15~and 5~are binary, triple and
quadruple, respectively, these being realistic proportions. There are
thus 85~companion stars which the observer is not aware of if none of
the multiple systems are resolved. Because the distribution of
secondary masses for a given primary mass is not uniform but typically
increases with decreasing mass \cite{MZ01}, the bias is such that
low-mass stars are underrepresented in any survey that does not detect
companions \cite{KTG91,Hetal98,L98,MZ01}.

Comprehensive star-count analysis of the solar neighborhood need to
incorporate unresolved binary systems, metalicity and age spreads and
the density fall-off perpendicular to the Galactic disk. Such studies
show that the IMF can be approximated by a two-part power-law with
$\alpha_1=1.3\pm0.7, 0.08 <m/M_\odot \le 0.5$ and $\alpha_2=2.2,
0.5<m/M_\odot \le 1$, a result obtained for two different MLRs
\cite{K01b}. Fig.~\ref{fig:lfmods} demonstrates simplified models
that, however, take into account a realistic population of triple and
quadruple stellar systems. The two best-fitting MLRs shown in
Fig.~\ref{fig:mlr} are used. The difference between the single-star
and system LFs is evident in all cases, being most of the explanation
of the disputed \cite{LFdiscr} discrepancy between the observed
$\Psi_{\rm near}$ and $\Psi_{\rm phot}$. It is also evident however,
that the model system LFs do not approximate $\Psi_{\rm phot}$ very
well. This is probably due to the used MLRs not accounting for the
full height of the maximum in the LF. 

\paragraph*{Star clusters}

Most star clusters offer populations that are co-eval and equidistant
with the same chemical composition. As a compensation for these
advantages the extraction of faint cluster members is very arduous
because of contamination from the background Galactic-field
population. The first step is to obtain photometry of everything
stellar in the vicinity of a cluster and to select only those stars
that lie near one or a range of isochrones, taking into account that
unresolved binaries are brighter than single stars. The next step is
to measure proper motions and radial velocities of all candidates to
select only those high-probability members that have coinciding space
motion with a dispersion consistent with the a priori unknown but
estimated internal kinematics of the cluster. Because nearby clusters
for which proper-motion measurements are possible appear large on the
sky, the observational effort is horrendous. For clusters such as
globulars that are isolated the second step can be omitted, but in
dense clusters stars missed due to crowding need to be corrected for.
The stellar LFs in clusters turn out to have the same general shape as
the photometric Galactic-field LF, $\Psi_{\rm phot}$
(Fig.~\ref{fig:lfs}), although the maximum is slightly offset
depending on the metalicity of the population \cite{KT97}. This
beautifully confirms that the maximum in the LF is due to structure in
the derivative of the MLR.  A 100~Ma isochrone (the age of the
Pleiades) is also plotted in Fig.~\ref{fig:mlr} to emphasize that for
young clusters additional structure in the LF is expected
(eq.~\ref{eq:mf_lf}).  This is due to stars with $m<0.6\,M_\odot$ not
having reached the main-sequence yet \cite{Belikov98,ChB00}.

LFs for star clusters are, like $\Psi_{\rm phot}$, system LFs because
binary systems are not resolved in the typical star-count survey. The
binary-star population evolves due to encounters. After a few initial
crossing times only those binary systems survive that have a binding
energy larger than the typical kinetic energy of stars in the
cluster. Calculations of the formation of an open star cluster
demonstrate that the binary properties of stars remaining in the
cluster are comparable to those in the Galactic field even if all
stars initially form in binary systems \cite{KAH}.  A further
disadvantage of cluster LFs is that star clusters preferentially loose
single low-mass stars across the tidal boundary as a result of
ever-continuing re-distribution of energy during encounters.  With
time, the retained population has an increasing binary proportion and
increasing average stellar mass. The global PDMF thus flattens with
time with a rate inversely proportional to the relaxation time. For
highly evolved initially rich open clusters it evolves towards a delta
function near the turnoff mass.

If a star cluster is younger than a few~Ma classical pre-main sequence
theory fails. This theory assumes hydrostatic contraction of spherical
non-rotating or sometimes slowly rotating stars from idealized initial
states.  However, Wuchterl has shown that stars this young remember
their accretion history \cite{WK01}. They are rotating rapidly and
are non-spherical. Pre-main sequence tracks taking these effects into
account are not available yet because of the severe computational
difficulties.  Estimates of the IMF in such very young clusters have
to resort to classical calculations despite this gap in our
theoretical understanding.  Furthermore, the age-spread of stars is
comparable to their age requiring spectroscopic classification of
individual stars to place them on a theoretical (but hitherto
classical) isochrone to estimate their masses \cite{Meyer00}.  Binary
systems are also not resolved.  A few results are shown in
Fig.~\ref{fig:mfn}. Taking the Orion nebula cluster (ONC) as the
best-studied example \cite{HC00,L00,MLL00}, the figure shows how the
shape of the deduced IMF varies with improving (but still classical)
pre-main sequence evolution calculations. This demonstrates that any
apparent sub-structure in the IMF cannot yet be relied upon to reflect
possible underlying physical mechanisms of star formation.

For the much more massive and long-lived globular clusters
($N\simgreat 10^5$~stars) theoretical stellar-dynamical work shows
that the MF measured for stars near the cluster's half-mass radius is
similar to the global PDMF. Inwards and outwards of this radius the MF
is flatter (smaller $\alpha$) and steeper (larger $\alpha$),
respectively. This comes from dynamical mass segregation
\cite{VH97}. Strong mass loss in a strong tidal field flattens the
global PDMF such that it no longer resembles the IMF anywhere
\cite{PZ99}.

\paragraph*{Brown dwarfs}

Brown dwarfs were theoretical constructs since the early 1960's
\cite{HN63} until the first cases were discovered in 1995
\cite{Basri00}.  For the solar neighborhood, near-infrared large-scale
surveys have now identified about~50 BDs probably closer than
25~pc. Because these objects do not have reliable distance
measurements an ambiguity exists between their ages and
distances. Only statistical analysis which relies on an assumed SFH
for the solar neighborhood can presently constrain the IMF, finding
$\alpha_0\simless 1$ for the Galactic-field BD IMF \cite{Ch01b}.

Surveys of young star clusters have also discovered BDs by finding
objects that extend the color--magnitude relation towards the faint
locus while being kinematical members. Given the great difficulty of
this endeavor only a few clusters now possess constraints on the
IMF. The Pleiades star cluster has proven especially useful, given its
proximity ($\approx127$~pc) and young age ($\approx100$~Ma). Results
indicate $\alpha_0\approx0.5-0.6$ (Table~\ref{tab:apl}). Estimates for
other clusters (ONC, $\sigma$~Ori, IC~348; Table~\ref{tab:apl}) also
indicate $\alpha_0\simless0.8$.

There appears to be no lower-mass limit for BDs. Free-floating planets
(FFLOPs) ($\simless 0.01\,M_\odot$) have been discovered in the very
young ONC \cite{Lucas00,Lucas01} and in the $\sigma$~Orionis cluster
\cite{Netal01,Netal01b,Betal01}. The IMF for FFLOPs appears to be
similar to that for the more massive BDs.

The above estimates of the IMF suffer under the same bias affecting
stars, namely unseen companions. BD--BD binary systems are known to
exist \cite{Basri00}, notably in the Pleiades cluster where their
offset in the color--magnitude diagram from the single-BD locus makes
them conspicuous. But their frequency is not yet very well constrained
because detailed scrutiny of individual objects is time-intensive on
large telescopes.  Calculations \cite{KAH,K01a} of the formation and
dynamical evolution of star clusters show that after a few crossing
times the binary proportion among BDs is smaller than among low-mass
stars. The distribution of separations does not extend to the same
distances as for stellar systems. This is a result of the weaker
binding energy of BD--BD binaries.  These calculations also show that
after a few crossing times the star--BD binary proportion is smaller
than the star--star binary proportion. This is consistent with the
results of a number of searches that have found no wide BD companions
to nearby stars \cite{Basri00}.  Radial-velocity surveys of BD
companions to nearby low-mass stars also show that star--BD binaries
are very rare for separations $\simless 3$~AU. The general absence of
BD companions is referred to as the {\it brown-dwarf desert}, because
stellar companions and planets are found at such separations
\cite{Halbw00,Vogt01}.  A few very wide systems with BD companions can
form during the final stages of dissolution of a small cluster
\cite{Fuente98}, and three such common proper-motion pairs have
perhaps been found \cite{Gizis01}.

\paragraph*{The average IMF}

The constraints arrived at above for $m\simless1\,M_\odot$ and
$m\simgreat 8\,M_\odot$ can be conveniently described by a multi-part
power-law form (eqs.~\ref{eq:imf_mult} and~\ref{eq:imf} in
Table~\ref{tab:imfs}).  Because this IMF has been obtained from
solar-neighborhood data for low-mass and very low-mass stars and from
many clusters and OB associations for massive stars it is an average
IMF. For $m<1\,M_\odot$ is is the IMF for single stars because unseen
companions are corrected for in this sample.  Independent measurements
of the IMF are consistent with the average multi-part power-law form
(Fig.~\ref{fig:apl}) .

The number fractions, mass fractions and mass densities contributed to
the Galactic-field total by stars in different mass-ranges are
summarized in Table~\ref{tab:frac}. Main-sequence stars make up about
half of the baryonic matter density in the local Galactic disk.  Of
the stellar contribution to the matter density, BDs make up about
40~\% in number and about 7~\% in mass.  The numbers in the table are
consistent with observed star-formation events such as in
Taurus--Auriga (TA). In TA groups of a few dozen stars form that do
not contain stars more massive than the Sun.  The table also shows
that a star cluster loses about 10~\% of its mass through stellar
evolution within 10~My if $\alpha_3=2.3$ (turnoff-mass $m_{\rm
to}\approx20\,M_\odot$), or within 300~My if $\alpha_3=2.7$
(turnoff-mass $m_{\rm to}\approx3\,M_\odot$). After about 10~Gy the
mass loss through stellar evolution alone amounts to about 40~\% if
$\alpha_3=2.3$ or 30~\% if $\alpha_3=2.7$.  Mass loss through stellar
evolution therefore poses no risk for the survival of star clusters
for the IMFs discussed here, because the mass-loss rate is small
enough for the cluster to adiabatically re-adjust.  A star-cluster may
be destroyed through mass loss from supernova explosions if
$\alpha\approx1.4$ for $8<m/\,M_\odot\le 120$ which would mean a
mass-loss of 50~\% within about 40~My when the last supernova explodes
\cite{K01a}.  None of the measurements in a resolved population has
found such a low $\alpha$ for massive stars (Fig.~\ref{fig:apl}).

\section*{Variation of the IMF and Theoretical Aspects} 

Is the scatter of data points in the alpha-plot (Fig.~\ref{fig:apl}) a
result of IMF variations? Before this can be answered affirmatively
any non-physical sources for scatter in the power-law index
determinations need to be assessed.

For a truly convincing departure from the average IMF a measurement
would need to lie outside the conservative uncertainty range of the
average IMF.  Significant departures from the average IMF only occur
in the shaded areas of the alpha plot. These are, however, not
reliable. The upper mass range in the shaded area near $1\,M_\odot$
poses the problem that the star-clusters have evolved such that the
turn-off mass is near to this range so that conversion to masses
critically depends on stellar-evolution theory and the adopted cluster
ages.  Some clusters such as $\rho$~Oph are so sparse that more
massive stars did not form. In both these cases the shaded range is
close to the upper mass limit. This leads to possible stochastic
stellar-dynamical biases because the most massive stars meet near the
core of a cluster due to mass segregation, but three-body or
higher-order encounters there can cause expulsions from the cluster.
The shaded area near $0.1\,M_\odot$ poses the problem that the
low-mass stars are not on the main sequence for most of the clusters
studied. They are also prone to bias through mass-segregation by being
underrepresented within the central cluster area that is easiest to
study observationally. Especially the latter is probably biasing the
M35 datum. Some effect with metalicity may be operating though,
because M35 appears to have a smaller $\alpha$ near the H-burning mass
limit than the Pleiades cluster which has a similar age but has a
larger abundance of metals (Fig.~\ref{fig:mfn}).

Measurements of the IMF for massive stars that are too far from
star-forming sites to have drifted to their positions within their
life-times yield $\alpha_3\approx4.5$ \cite{M98}. This value is
discordant with the average IMF and is often quoted to be a good
example of evidence for a varying IMF, being the result of isolated
high-mass star-formation in small clouds. However, accurate
proper-motion measurements show that even the firmest members of this
isolated population have very high space motions \cite{Ram01}. Such
high velocities are most probably the result of energetic
stellar-dynamical ejections when massive binary systems interact in
the cores of star-clusters in normal but intense star-forming regions
located in the MW disk. The large $\alpha_3$ then probably comes about
because the typical ejection velocity is a decreasing function of
ejected stellar mass, but detailed theoretical verification is not yet
available.

To address such stellar-dynamical biases an extensive theoretical
library of binary-rich star clusters has been assembled \cite{K01a}
covering 150~My of stellar-dynamical evolution taking into account
stellar evolution and assuming the average IMF in all cases.
Evaluating the MF within and outside of the clusters, at different
times and for clusters containing initially $800-10^4$ stars leads to
a theoretical alpha-plot which reproduces the spread in $\alpha(lm)$
values evident in the empirical alpha-plot (Fig.~\ref{fig:apl}). This
verifies the conservative uncertainties adopted in the average IMF but
implies that the scatter in the empirical alpha-plot around the
average IMF cannot be interpreted as true variations.

Enough IMF data have been compiled to attempt the first analysis of
the distribution of power-law indices. If all stellar populations have
the same IMF then this should be reflected by this distribution.  It
ought to be a Gaussian with a mean $<\!\!\alpha\!\!>$ value
corresponding to the true IMF, and a dispersion reflecting the
measurement uncertainties.  The distribution of $\alpha$ data for
$m>2.5\,M_\odot$ (Fig.~\ref{fig:ahist}) shows a narrow peak positioned
at the Salpeter value, with symmetric broad wings.  The empirical data
are therefore not distributed like a single Gaussian function.  The
theoretical alpha-plot shows a distribution consistent with a single
Gaussian.  Its width is comparable to the broad wings in the empirical
data.  Interestingly, the spread, $\sigma_{\alpha,f}=0.08$, of the
narrow peak in the empirical data is very similar to the uncertainties
quoted by Massey in an extensive observational determination of the
IMF for massive stars, $\alpha=2.2\pm0.1$. It is not clear at this
stage if the empirical distribution does reflect true IMF variations.
The symmetry of the broad wings suggests a superposition of at least
two Gaussians with different measurement uncertainties but the same
underlying IMF for massive stars.

If $\alpha_3= 2.3\pm0.1$ is adopted for massive stars, then the
measurement $\alpha=1.6\pm0.1$ for the massive Arches cluster
(Table~\ref{tab:apl}), which is situated near the Galactic center and
difficult to observe, would definitely mean an IMF that is top-heavy
for this extreme population.  There are also indications of top-heavy
IMFs in star clusters in the starburst \cite{starburst} galaxy~M82
which has a low metalicity. The galaxy is too distant for its clusters
to be resolved into individual stars and binaries, so that the stellar
LF cannot be measured. However, spectroscopy of the massive M82-F
cluster allows measurement of the velocity dispersion of the stars in
the cluster. Together with the cluster size this gives a mass for the
cluster if it is assumed that the cluster is in gravitational
equilibrium. The derived mass-to-light ratio is significantly smaller
than the ratio expected from the average IMF for such a young (about
60~Ma) population. The implication is that the M82-F population is
significantly depleted in low-mass stars, or top-heavy \cite{SG01}.
Stellar-dynamical modeling of forming star clusters is needed to
investigate if M82-F may have been stripped off its low-mass stars by
the tidal field. Furthermore, X-ray observations of M82 suggest that
the relative abundances of some heavy elements seem to be inconsistent
with the expectation of the Salpeter IMF, and that stars with masses
above $25\,M_\odot$ seem to contribute significantly to the metal
enrichment of the galaxy \cite{Tsuru,Umeda}.  These studies are
independent of the unresolved cluster issue and suggest that the slope
of the IMF for massive stars is likely to be smaller than the Salpeter
value, $\alpha_3\simless 2$. This indirect approach, however, relies
on exact knowledge of nucleosynthesis yields and the processes
governing injection of enriched material back into the ISM.
Additional evidence for variations of the IMF come from the central
regions of very young star clusters. For example, the center of the
ONC is deficient in low-mass stars (Fig.~\ref{fig:mfn}) although the
global MF for this cluster is similar to the average IMF.  The
interpretation of a locally varying IMF depends on whether mass
segregation in the ONC is primordial, or whether it is the result of
stellar-dynamical evolution.  

Two well-studied and resolved starburst clusters have $\alpha_3
\approx\,2.3$ (30~Dor and NGC~3603, Table~\ref{tab:apl}). These are
also massive and very young clusters, but they oppose the suggestion
from the Arches and M82-F clusters that starbursts may prefer
top-heavy IMFs.  From the ONC we know that the entire mass spectrum
$0.05\simless m/M_\odot\simless 60$ is present roughly following the
average IMF (Fig.~\ref{fig:mfn}).  Low-mass stars are also known to
form in the much more massive 30~Dor cluster \cite{Setal00} although
their IMF has not been measured yet due to the large distance of about
55~kpc.  The available evidence is thus that low-mass stars and
massive stars form together even in extreme environments without, as
yet, convincing demonstration of a variation of the number ratio.

The observational study by Luhman \cite{L00} of many close-by
star-forming regions using one consistent methodology finds that the
IMF does not show measurable differences from low-density star-forming
regions in small molecular clouds ($n= 0.2-1$~stars/pc$^3$ in
$\rho$~Oph) to high-density cases in giant molecular clouds ($n=
(1-5)\times 10^4$~stars/pc$^3$ in the ONC). This result extends to the
populations in the truly exotic ancient and metal-poor
dwarf-spheroidal satellite galaxies. These are speculated to be
dominated by dark matter and thus probably formed under conditions
that were different from present-day events.  Two such close
companions to the MW have been observed \cite{Grill98,Felt99} finding
the same MF as in globular clusters for $0.5\simless m/M_\odot
\simless 0.9$. Thus, again there are no significant differences to the
average IMF. This apparent universality of the IMF is also supported
by available chemical evolution models of the MW \cite{Chiap}.  The
IMF for metal-poor and metal-rich populations of massive stars is the
same \cite{M98}.  Between about $10\,M_\odot$ and $m_{\rm
u}>70-100\,M_\odot$ the IMF is a power-law with $\alpha=2.1\pm0.1$ for
13~clusters and OB associations in the MW (metalicity $Z \approx 0.02
= Z_\odot$, which is the Solar mass fraction of metals),
$\alpha=2.3\pm0.1$ for 10~clusters and OB associations in the Large
Magellanic Cloud ($Z=0.008$) and $\alpha=2.3\pm0.1$ for one cluster in
the Small Magellanic Cloud ($Z=0.002$). The data imply that the mass
of the most massive star, $m_{\rm max}>70-100\,M_\odot$, is
independent of $Z$, and only depends on the number of stars in the
star-forming event. The most massive star that is present in a
population is consistent with stars being sampled randomly from the
IMF without an upper mass limit, $m_{\rm max}$, the IMF taking on the
meaning of a probability density function.  This questions the concept
of a fundamental maximum upper stellar mass, although unresolved
multiple systems may be mistaken for very massive stars. It follows
that radiation pressure on dust grains during star-assembly cannot be
a physical mechanism establishing $m_{\rm max}$ \cite{assembly}.

However, there may be some IMF variation for very-low mass stars.
Present-day star-forming clouds typically have somewhat higher
metal-abundances (log$_{10}(Z/Z_\odot) \approx\;$
[Fe/H]$\;\approx+0.2$) compared to 6~Ga ago ([Fe/H] $\;\approx-0.3$)
\cite{BM98}. This is the mean age of the population defining the
average IMF. The data in the empirical alpha-plot indicate that some
of the younger clusters may have a single-star IMF that is somewhat
steeper than the average IMF if unresolved binary-stars are corrected
for \cite{K01a}. Clouds with a larger [Fe/H] appear to produce
relatively more very low-mass stars.  This is tentatively supported by
the M35 result (Fig.~\ref{fig:mfn}) and by the typically flatter MFs
in globular clusters \cite{PZ99} that have [Fe/H]$\;\approx -1.5$. The
recent finding that the old and metal-poor ([Fe/H] $\;\approx-0.6$)
thick-disk population has a flatter IMF below $0.3\,M_\odot$ with
$\alpha\approx0.5$ \cite{RR01} also supports this assertion.  If such
a systematic effect is present, then for $m\simless 0.7\,M_\odot$,
\begin{equation}
\alpha \approx 1.3 + \Delta\alpha\,{\rm [Fe/H]},
\end{equation}
with $\Delta\alpha \approx0.5$.  Many IMF measurements are needed to
verify if such a variation exists because it is within the present
uncertainty in $\alpha$. As a possible counterexample, the IMF
measured for spheroidal MW stars that have [Fe/H]$\;\approx-1.5$ does
not appear to be significantly flatter than the average IMF
\cite{GFB}, so the issue is far from being settled.

Theoretical considerations do suggest that for sufficiently small
metalicity a gas cloud cannot cool efficiently causing the Jeans mass
required for gravitational collapse to be larger. In particular, the
first stars ought to have large masses because of this effect
\cite{Larson98,Bromm}. If the IMF of the first stars were similar to
the average IMF then long-lived low-mass stars should exist that have
no metals. However, none have been found \cite{Beers}, possibly
implying that the IMF of the first stars was very different from the
average IMF.  Finding the remnants of these first stars poses a major
challenge. An easier target is measuring the IMF for low-mass and
very-low mass stars in metal-poor environments, such as young
star-clusters in the Small Magellanic Cloud. Metalicity does play a
role in the planetary-mass regime because the detected exo-planets
occur mostly around stars that are more metal-rich than the Sun
\cite{Santos}.  This suggests that metal-richer environments may favor
the formation of less-massive objects.

While the Jeans-mass argument should be valid as a general indication
of the rough mass scale where fragmentation of a contracting gas cloud
occurs, the concept breaks down when considering the stellar masses
that form in star clusters. The central regions of clusters are
denser, formally leading to smaller Jeans masses which is the opposite
of the observed trend. Even in very young clusters massive stars tend
to be located in the inner regions. More complex physics is
involved. Stars may regulate their own mass by powerful outflows
\cite{AL96}, and the coagulation of protostars probably plays a role
in the densest regions where the cloud-core collapse time, $\tau_{\rm
coll}$, is longer than the fragment collision time-scale which is the
cluster crossing time, $t_{\rm cr}$. The collapse of a fragment to a
protostar with $\simgreat 90$~\% of the final stellar mass takes no
longer than $\tau_{\rm coll}\approx 0.1$~My \cite{WK01}, so that
$t_{\rm cr} <0.1$~My implies $M/R^3 > 10^5\, M_\odot$/pc$^{-3}$. Such
densities are only found in the centers of very populous embedded star
clusters. This may explain why massive stars are usually centrally
concentrated in very young clusters \cite{Bonn98,Kl01}.  However,
until accurate $N-$body computations are performed for a number of
cases, the observed mass segregation in very young clusters cannot be
taken as evidence for primordial mass segregation, and thus for
coagulation and local IMF variations.  For example, models of the ONC
show that the degree of observed mass segregation can be established
dynamically within about 2~My (Fig.~\ref{fig:mfn}) despite the
embedded and much denser configuration having no initial mass
segregation.

The origin of most stellar masses is indicated by recent observations
of star formation in the $\rho$~Oph cluster.  In this modest
proto-cluster the pre-stellar and protostar MFs are indistinguishable.
Both are indistinguishable from the average IMF upon correction for
binaries that presumably form in the cores \cite{Motte,Bont01}. The
pre-stellar cores have sizes and densities that agree with the
Jeans-instability argument for the conditions in the $\rho$~Oph cloud.
Cloud-fragmentation therefore appears to be the most-important
mechanism shaping the stellar IMF for masses $0.05\simless
m/M_\odot\simless\,3$, and the shape of the IMF is determined by the
spectrum of density fluctuations in the molecular cloud. The
computations of cloud fragmentation by Klessen are beginning to
reproduce the initial stages of this process \cite{Kl01b}, but suggest
that the emerging IMF depends on the star formation conditions.  The
empirical data indicate that stars freeze out of the molecular gas
much faster than the motions between the stars thereby preserving the
distribution of density fluctuations in the cloud \cite{Elm00}.  The
majority of stellar masses making up the average IMF thus do not
appear to suffer subsequent modifications such as competitive
accretion \cite{Bonn01b} or protostellar mergers.  In particular, the
flattening of the IMF near $0.5\,M_\odot$ does not appear to be a
result of the decay of few-body systems that eject unfinished
protostellar cores \cite{Reipurth01}, although this mechanism must
operate in at least some cases.  This notion as the dominant source of
BDs is also in conflict with the apparent abundance of BDs in the ONC
but the virtual absence of BDs within the TA star-forming clouds
\cite{Luhman}.  The ejection process should operate in both
environments.  The problem with the unfinished-protostellar-core
ejection scenario is that the BDs leave their parent cluster within a
time shorter than the cluster crossing time thus rendering them
unlikely to be seen in the cluster \cite{Bonn01}. However, the four
BDs detected in far-outlying regions of TA \cite{Martin} may
constitute examples of ejected cores.  The intriguing result from
$\rho$~Oph is consistent with the independent finding that the
properties of binary systems in the Galactic field can be understood
if most stars formed in modest $\rho$~Oph-type clusters with
primordial binary properties as observed in TA \cite{K95d}. However,
the average IMF is also similar to the MF in the dense ONC
(Fig.~\ref{fig:mfn}), implying that fragmentation of the pre-cluster
cloud there must have proceeded similarly. It is not clear why the
spectrum of density fluctuations in the pre-cluster cloud should have
been similar under such different conditions.


In summary, the Galactic-field IMF (eq.~\ref{eq:imf} in
Table~\ref{tab:imfs}) appears to be remarkably universal, with the
exception in the sub-stellar mass regime.  A weak empirical trend with
metalicity is suggested for very-low mass stars: More metal-rich
environments may be producing relatively more low-mass objects. For
massive stars a correlation with star-forming conditions has not been
found despite intense searches.  The evidence for top-heavy IMFs come
either from clusters that cannot be resolved or clusters that are very
difficult to observe, or from entirely indirect arguments such as
peculiar abundances of elements. This may mean that only in those rare
starburst cases that are not easily accessible to the observer does
the IMF begin to deviate towards a top-heavy form. Alternatively,
maybe presently not understood biases are affecting the interpretation
of such extreme systems that require indirect deductions about the
IMF.

Uncertainties of the IMF arise because of the bias due to unresolved
multiple systems and due to uncertainties in theoretical stellar
models with rotation and theoretical models for ages younger than
approximately one~Ma. For massive stars the true IMF may be closer to
Scalo's value $\alpha_3\approx2.7$ rather than the Salpeter value
$\alpha_3\approx2.3$.  This is valid for all studied populations
provided they have similar binary-star properties.  

The majority of stellar masses appear to be determined by the
fragmentation of molecular clouds with little subsequent modifications
such as ejections of unfinished cores or competitive accretion.  It is
unclear why this fragmentation process should lead to
indistinguishable IMFs despite very different star forming conditions.
There appears to be no empirical maximum stellar mass, nor an
empirical minimum mass for BDs.  Only for massive stars are cloud-core
or protostellar interactions probably important. BDs are probably
cores that lost their envelopes due to chance proximity to an O~star.
This hypothesis may explain their occurrence in relatively rich star
clusters and their virtual absence in TA.




\begin{thebibliography}{9}
\bibitem{kepler} Stellar masses can be measured directly in binary
systems. Unfortunately the Kepler orbits are available only for very
few well-studied cases. These do not constitute a volume-limited
unbiased sample.
\bibitem{magn} The absolute magnitude in the broad wavelength band $P$
is $M_P \equiv -2.5\,{\rm log}_{10}(L_P/L_{P,\odot}) + M_{P,0}$, where
$L_P, L_{P,\odot}$ are the stellar and Solar luminosities,
respectively, and $M_{P,0}$ is an empirical zero-point. The apparent
magnitude $m_P = M_P + 5\,{\rm log}_{10}d - 5$, where $d$ is the
distance of the star in pc.
\bibitem{field} Galactic field stars are not members of star clusters
but orbit in the potential of the Milky Way disk. The
solar-neighborhood \cite{neighb} constitutes a local field sample.
\bibitem{LFphot} The disadvantage of the LF created using this
technique is that the distance measurements are indirect by relying on
photometric parallax. The underlying principle of this technique is
to estimate the luminosity of a star from the stellar type obtained
from the color of the star or from spectral analysis. The apparent
luminosity then gives a distance estimate.  Such surveys are flux
limited rather than volume limited. Pencil-beam surveys which do not
pass through virtually the entire stellar disk are prone to Malmquist
bias. This bias results from a spread of luminosities of stars that
have the same color because of their dispersion of metalicities and
ages.  Intrinsically more luminous stars enter the flux-limited
sample.  This biases the inferred absolute luminosities and the
inferred stellar spatial densities.  Malmquist bias can be corrected
for \cite{SIP}, and is similar to the Lutz-Kelker bias \cite{LK}.
\bibitem{LFdiscr} The discrepancy evident in Fig.~\ref{fig:lfs}
between the nearby LF, $\Psi_{\rm near}$, and the photometric LF,
$\Psi_{\rm phot}$, invoked a significant dispute \cite{K95a,RG97} as
to the nature of this discrepancy. On the one hand \cite{K95a} the
difference is thought to be due to unseen companions in the deep but
low-resolution surveys used to construct $\Psi_{\rm phot}$, with the
possibility that photometric calibration for very-low-mass stars may
remain problematical so that the exact shape of $\Psi_{\rm phot}$ for
$M_V\simgreat 14$ is probably uncertain.  On the other hand
\cite{RG97} the difference is thought to come from non-linearities in
the $V-I, M_V$ color--magnitude relation used for photometric
parallax. Taking into account such structure it can be shown that the
photometric surveys underestimate stellar space densities so that
$\Psi_{\rm phot}$ moves closer to the extended estimate of $\Psi_{\rm
near}$ using a sample of stars within 8~pc or further.  While this is
an important point, the extended $\Psi_{\rm near}$ is incomplete
\cite{compl} and theoretical color-magnitude relations do not have
the required degree of non-linearity. The observational
color--magnitude data also do not conclusively suggest a feature with
the required strength \cite{BCAH98}. Furthermore, $\Psi_{\rm phot}$
agrees almost perfectly with the LFs measured for star clusters of
solar and population~II metalicity (Fig.~\ref{fig:lfs}) so that it
appears unlikely that non-linearities in the color--magnitude
relation significantly affect $\Psi_{\rm phot}$.
\bibitem{starburst} A starburst is a region in which star formation is
ongoing with a very high rate. For a star-formation event to be
classified as a starburst $> 10^6\,M_\odot$ of stars have to
be produced within about 1~My giving a star-formation rate of
$> 1\,M_\odot/$My \cite{IPI}.
\bibitem{assembly} The physics of formation of massive stars is 
controversial \cite{Duchene01}.  Radiation pressure from the growing
core of a massive star should halt spherical accretion for $m\simgreat
10\,M_\odot$ so that massive stars should be produced through
collisions of intermediate-mass protostars in dense cluster cores
\cite{Bonn98}. Accretion through massive disks or with very high rates
($\simgreat 10^{-5}\,M_\odot$/yr) may, however, overcome radiation
pressure causing the formation of massive stars to remain an unsolved
problem \cite{MB01}.  Possible physics limiting the mass of the most
massive stars is reviewed in \cite{Sc78}.  Very massive stars may
finish their main-sequence life before the accretion process completes
possibly rendering the most massive stars undetectable.
\bibitem{compl} Very low-mass stars cannot be detected to large
distances so that the nearby LF is poorly constrained.  It is
therefore important to increase the sample of nearby stars, but
controversy exists as to the maximum distance to which the
very-low-mass star census is complete. Using spectroscopic parallax it
has been suggested that the local census of very-low mass stars is
complete to distances of 8~pc and beyond \cite{RG97}.  However,
Malmquist bias \cite{LFphot} allows stars and unresolved binaries to
enter such a flux-limited sample from much larger distances
\cite{K01b}.  The increase of the number of stars with distance using
trigonometric distance measurements shows that the nearby sample
becomes severely incomplete for distances larger than 5~pc and for
$M_V>12$ \cite{J94,Henry97}. Recently discovered companions
\cite{Detal99,Beuzit01} to known primaries in the distance range
$5<d<12$~pc confirm that the extended sample is not yet complete.
\bibitem{LK} The Lutz-Kelker bias affects volume-limited star-count
surveys in a similar way as the Malmquist bias \cite{LFphot} affects
flux-limited surveys. Distance observations have errors. Because the
number of stars increases non-linearly with the distance there are
more stars just outside the formal distance limit than inside.  There
are thus more stars with true distances outside the limit that enter
as a result of erroneous measurements than stars that are discounted
from the survey because erroneous measurements place them outside
although they are actually nearer than the distance limit. This
results in a bias in the deduced average distances and stellar number
densities \cite{LuKe}.
\end{thebibliography}

\bibliographystyle{Science}


\begin{scilastnote}
\item I acknowledge use of the NASA Astrophysics Data System and
partial support through DFG grant KR1635. I thank Lynne Hillenbrand,
Xavier Delfosse and David Barrado Y Navascues for making available
their data. I am very grateful to Linda Rowan, Philippe Andr\'e,
Gilles Chabrier, Christopher Tout, Volker Weidemann, Karsten Weidner
and G\"unther Wuchterl for constructive comments.
\end{scilastnote}


\pagestyle{empty}

\begin{table}
{\small
\begin{tabular}{l|l|l|l}

\hline\hline
general
&\multicolumn{2}{l}{$dN = \xi(m)\,dm = \xi_{\rm L}(m)dlm$}\\
&\multicolumn{2}{l}{$\xi_{\rm L}(m) = (m\,{\rm ln}10)\,\xi(m)$}
&{\it gen}\\

Scalo's IMF index \cite{Sc86} 
&\multicolumn{2}{l}{$\Gamma(m)\equiv {d\over dlm}
\left({\rm log}_{10}\xi_{\rm L}(lm)\right)$}
&{\it Gam}\\

&\multicolumn{2}{l}{$\Gamma = -x = 1+\gamma = 1-\alpha$}
&{\it ind}\\
&e.g.  for power-law form:
&$\xi_{\rm L} = A\,m^\Gamma = A\,m^{-x}$\\
&&$\xi = A'\,m^\alpha = A'\,m^{-\gamma}$\\
&&$A' = A/{\rm ln}10$\\

\hline\hline

Salpeter(1955) \cite{S55} 
&$\xi_{\rm L}(lm) = A\,m^\Gamma$
&$\Gamma=-1.35 \, (\alpha=2.35)$
&{\it S}\\

&\multicolumn{2}{l}{$A = 0.03\,{\rm pc}^{-3}\,{\rm
log}_{10}^{-1}M_\odot; \quad 0.4\le m/M_\odot \le 10$}\\

\hline

Miller-Scalo(1979) \cite{MS79} 
&$\xi_{\rm L}(lm) = A\,{\rm exp}\left[-{\left(lm-lm_o\right)^2 \over 
2\,\sigma_{lm}^2 } \right]$
&$\Gamma(lm) = -{\left(lm-lm_o\right) \over \sigma_{lm}^2}\,{\rm log}_{10}e$
&{\it MS}\\

{\it thick long-dash-dotted line}
&\multicolumn{2}{l}{$A = 
106\,{\rm pc}^{-2}\,{\rm log}_{10}^{-1}M_\odot;
\quad lm_o=-1.02; \quad \sigma_{lm}=0.68$}\\

\hline

Larson(1998) \cite{Larson98}
&$\xi_{\rm L}(lm) = A\, m^{-1.35} {\rm exp}\left[-{m_o\over
                    m}\right]$
&$\Gamma(lm) = -1.35  + {m_o\over m}$
&{\it La}\\
{\it thin short-dashed line}
&\multicolumn{2}{l}{$A=-\,; \quad\quad m_o=0.3\,M_\odot$}\\

\hline

Larson(1998) \cite{Larson98}
&$\xi_{\rm L}(lm) = A\,\left[1 + {m\over m_o}\right]^{-1.35}$ 
&$\Gamma(lm) = -1.35\left(1 + {m_o\over m}\right)^{-1}$
&{\it Lb}\\
{\it thin long-dashed line}
&\multicolumn{2}{l}{$A=-\,; \quad\quad m_o=1\,M_\odot$}\\

\hline

Chabrier(2001) \cite{Ch01a,Ch01b}
&$\xi(m) = A\,m^{-\delta}\,{\rm exp}
         \left[-\left({m_o\over m}\right)^\beta \right]$
&$\Gamma(lm) = 1 - \delta + \beta\left(m_o\over m\right)^\beta$
&{\it Ch}\\

{\it thick short-dash-dotted line}
&\multicolumn{2}{l}{$A=3.0\,{\rm pc}^{-3}\,M_\odot^{-1};
\quad m_o=716.4\,M_\odot; \quad \delta=3.3;  \quad
\beta=0.25 $}\\

\hline

\end{tabular}
}{fig:apl}

{\small The multi-part power-law IMF:}
\begin{equation}
\xi (m) = k\left\{
          \begin{array}{l@{\quad\quad,\quad}l@{\quad,\quad}l}
   \left({m\over m_1}\right)^{-\alpha_0}  &m_0 < m \le m_1 &n=0\\
   \left({m\over m_1}\right)^{-\alpha_1}  &m_1 < m \le m_2 &n=1\\
   \left[
       \prod\limits^{n\ge2}_{i=2}\left({m_i\over
          m_{i-1}}\right)^{-\alpha_{i-1}}
       \right] 
        \left({m\over m_n}\right)^{-\alpha_n} 
        &m_n < m \le m_{n+1}  &n\ge2\\
          \end{array}\right.
\label{eq:imf_mult}
\end{equation}
{\small The average, or Galactic-field, single-star IMF has
$k=0.877\pm0.045$~stars/(pc$^3\,M_\odot$) for scaling to the solar
neighborhood with }
\begin{equation}
          \begin{array}{l@{\quad\quad,\quad}ll@{\quad,\quad}l}
\alpha_0 = +0.3\pm0.7   &0.01 &\le m/M_\odot < 0.08 &n=0 \\
\alpha_1 = +1.3\pm0.5   &0.08 &\le m/M_\odot < 0.50 &n=1 \\
\alpha_2 = +2.3\pm0.3   &0.5  &\le m/M_\odot < 1    &n=2 \\
\alpha_3 = _{+2.3\pm0.3}^{+2.7\pm0.3} &1 &\le m/M_\odot   &n=3.\\
          \end{array}
\label{eq:imf}
\end{equation}

\hrulefill

\vspace{-4mm}

\hrulefill

\caption{\small{Summary of different proposed analytical IMF forms.
Notation: $lm\equiv{\rm log}_{10}(m/M_\odot)={\rm ln}(m/M_\odot)/{\rm
ln}10$; $dN$ is the number of single stars in the mass interval $m$ to
$m+dm$ and in the logarithmic-mass interval $lm$ to $lm+dlm$.  The
mass-dependent IMF indices, $\alpha(m)$ (eq.~{\it ind}), are plotted
in Fig.~\ref{fig:apl} using the line-types defined here.  Eq.~{\it MS}
was derived by Miller\&Scalo assuming a constant star-formation rate
and a Galactic disk age of 12~Ga (the uncertainty of which is
indicated in the lower panel of Fig.~\ref{fig:apl}). Larson
\cite{Larson98} does not fit his forms (eqs.~{\it La} and~{\it Lb}) to
solar-neighborhood star-count data but rather uses these to discuss
general aspects of likely systematic IMF evolution; the $m_o$ in
eq.~{\it La} and~{\it Lb} given here are approximate eye-ball fits to
the average IMF. In the multi-power-law IMF, $\alpha_3=2.3$ is
consistent with the data (Fig.~\ref{fig:apl}), but correction for
unresolved binary systems increases this to $\alpha_3=2.7$.  The
uncertainties correspond to a 99~\% confidence interval for
$m>0.5\,M_\odot$ (Fig.~\ref{fig:apl}), and to a 95~\% confidence
interval for $0.1-0.5\,M_\odot$ \cite{KTG93}.  The nearby Hipparcos
LF, $\Psi_{\rm near}({\rm Hipp})$ (Fig.~\ref{fig:lfs}), has
$\rho=(5.9\pm0.3)\times10^{-3}$~stars/pc$^3$ in the interval
$M_V=5.5-7.5$ corresponding to the mass interval $m_2=0.891 -
0.687\,M_\odot$ \cite{K01b} using the KTG93 MLR
(Fig.~\ref{fig:mlr}). $\int_{m_1}^{m_2}\xi(m)\,dm=\rho$ yields
$k=0.877\pm0.045$~stars/(pc$^3\,M_\odot$).  }}
\label{tab:imfs}
\end{table}


\begin{table}
{\small
\begin{tabular}{l|ccc|ccc|c|c}

\hline\hline

mass range     
&\multicolumn{3}{c|}{$\eta_N$}
&\multicolumn{3}{c|}{$\eta_M$}    
&$\rho^{\rm st}$
&$\Sigma^{\rm st}$  \\

[$M_\odot$]    
&\multicolumn{3}{c|}{[per cent]}
&\multicolumn{3}{c|}{[per cent]}
&[$M_\odot/{\rm pc}^3$]  
&[$M_\odot/{\rm pc}^2$] \\

&\multicolumn{3}{c|}{$\alpha_3$}
&\multicolumn{3}{c|}{$\alpha_3$}    
&$\alpha_3$
&$\alpha_3$  \\

&2.3 &2.7 &4.5 &2.3 &2.7 &4.5 &4.5 &4.5\\

\hline

0.01--0.08 
&37
&38 
&39

&4.1
&5.4
&7.4

&$3.2\times10^{-3}$
&1.6

\\ 

0.08--0.5
&48
&49
&50

&27
&35
&48

&$2.1\times10^{-2}$
&10

\\ 

0.5--1
&8.9
&9.1
&9.3

&16
&21
&29

&$1.3\times10^{-2}$
&6.4

\\ 

1 -- 8
&5.7
&4.6
&2.4

&32
&30
&15

&$6.5\times10^{-3}$
&1.2

\\ 

8 -- 120 
&0.40
&0.14
&0.00

&21
&7.8
&0.08

&$3.6\times10^{-5}$
&$6.5\times10^{-3}$

\\ 

\hline

$\overline{m}/M_\odot=$
&$0.38$
&$0.29$
&$0.22$

&
&
&

&$\rho_{\rm tot}^{\rm st}=0.043$
&$\Sigma_{\rm tot}^{\rm st}=19.6$

\\

\hline \hline

&\multicolumn{1}{r|}{}
&\multicolumn{2}{c|}{$\alpha_3=2.3$} 
&\multicolumn{2}{c||}{$\alpha_3=2.7$}
&
&\multicolumn{2}{c}{$\Delta M_{\rm cl}/M_{\rm cl}$}
\\ 

&\multicolumn{1}{r|}{$m_{\rm max}$} 
&$N_{\rm cl}$  &$M_{\rm cl}$  
&$N_{\rm cl}$  &\multicolumn{1}{r||}{$M_{\rm cl}$}
&$m_{\rm to}$  &\multicolumn{2}{c}{[per cent]}
\\

&\multicolumn{1}{r|}{[$M_\odot$]} & &[$M_\odot$] & 
&\multicolumn{1}{r||}{[$M_\odot$]} &[$M_\odot$] 
&$\alpha_3=2.3$ &$\alpha_3=2.7$\\

\cline{2-9}

&\multicolumn{1}{r|}{1}      &16    &2.9     &21                 
&\multicolumn{1}{r||}{3.8}
&80 &2.1 &0.5
\\
								     
&\multicolumn{1}{r|}{8}      &250   &74     &730                
&\multicolumn{1}{r||}{200}
&60 &3.8 &0.9
\\
								      
&\multicolumn{1}{r|}{20}     &810   &270     &3400               
&\multicolumn{1}{r||}{970}
&40 &6.5 &1.6
\\
								      
&\multicolumn{1}{r|}{40}     &2000  &700     &$1.1\times10^4$    
&\multicolumn{1}{r||}{2300}
&20 &12 &3.5
\\

&\multicolumn{1}{r|}{60}     &3400 &1200     &$2.2\times10^4$    
&\multicolumn{1}{r||}{6400}
&8 &21 &7.8
\\

&\multicolumn{1}{r|}{80}     &4900 &1800     &$3.6\times10^4$    
&\multicolumn{1}{r||}{$1.1\times10^4$}
&3 &24 &9.7
\\
  
&\multicolumn{1}{r|}{100}    &6500 &2500     &$5.3\times10^4$    
&\multicolumn{1}{r||}{$1.5\times10^4$}
&1 &36 &24
\\

&\multicolumn{1}{r|}{120}    &8300 &3100     &$7.2\times10^4$    
&\multicolumn{1}{r||}{$2.1\times10^4$}
&0.7 &39 &28
\\

\hline\hline

\end{tabular}
}
\caption{\small{The number fraction is $\eta_N=100\,\int_{m_1}^{m_2}
\xi(m)\,dm/ \int_{m_l}^{m_u}\xi(m)\,dm$. The mass fraction is
$\eta_M=100\,\int_{m_1}^{m_2} m\,\xi(m)\,dm/ M_{\rm cl}$, $M_{\rm cl}=
\int_{m_l}^{m_u} m\,\xi(m)\,dm$. Both are in per cent for
main-sequence stars in mass intervals $m_1$ to $m_2$.  The stellar
contribution to the Oort limit, $\rho^{\rm st}$, and to the
Galactic-disk surface mass-density, $\Sigma^{\rm st}=2\,h\rho^{\rm
st}$. The above quantities assume for the lower and upper mass limits,
respectively, $m_l=0.01\,M_\odot$ and $m_u=120\,M_\odot$. The
Galactic-disk scale-height $h=250$~pc for $m<1\,M_\odot$ \cite{KTG93}
and $h=90$~pc for $m>1\,M_\odot$ \cite{Sc86}. Results are shown for
the average IMF (eq.~\ref{eq:imf} in Table~\ref{tab:imfs}), for the
high-mass-star IMF approximately corrected for unresolved companions
($\alpha_3=2.7, m>1\,M_\odot$), and for the PDMF in the solar
neighborhood ($\alpha_3=4.5$ \cite{Sc86,KTG93}) which describes the
distribution of stellar masses now populating the Galactic disk. The
ISM contributes $\Sigma^{\rm ISM}=13\pm3\,M_\odot$/pc$^2$, $\rho^{\rm
ISM}\approx0.04\pm0.02\,M_\odot$/pc$^3$ and stellar remnants
contribute $\Sigma^{\rm rem}\approx3\,M_\odot$/pc$^2$, $\rho^{\rm
rem}\approx0.003\,M_\odot$/pc$^3$ \cite{Weide}.  BDs do not constitute
a dynamically important mass component of the Galaxy, even when
eq.~\ref{eq:imf} is extrapolated to $0.0\,M_\odot$ giving $\rho^{\rm
BD} = 3.3\times 10^{-3}\, M_\odot$/pc$^3$.  The average stellar mass
is $\overline{m}= \int_{m_l}^{m_u} m\,\xi(m)\,dm/
\int_{m_l}^{m_u}\xi(m)\,dm$.  $N_{\rm cl}$ is the number of stars that
have to form in a star cluster so that the most massive star in the
population has the mass $m_{\rm max}$. The mass of this population is
$M_{\rm cl}$, and the condition is $\int_{m_{\rm
max}}^{\infty}\xi(m)\,dm=1$ with $\int_{0.01}^{m_{\rm max}} \xi(m)\,dm
= N_{\rm cl}-1$. $\Delta M_{\rm cl}/M_{\rm cl}$ is the fraction of
mass lost from the cluster due to stellar evolution, assuming that for
$m\ge8\,M_\odot$ all neutron stars and black holes are kicked out due
to an asymmetrical supernova explosion, but that white dwarfs are
retained \cite{Wetal92} and have masses $m_{\rm WD} = 0.7\,M_\odot$
for progenitor masses $1\le m/M_\odot < 8$ and $m_{\rm
WD}=0.5\,M_\odot$ for $0.7\le m/M_\odot <1$. The evolution times for a
star of mass $m_{\rm to}$ to reach the turn-off age are available in
Fig.~\ref{fig:apl}.  }}
\label{tab:frac}
\end{table}


\begin{table}
{\small
\begin{tabular}{lccc}

\hline

&$\alpha$   &$\alpha$   &$\alpha$\\
&mass range [$M_\odot$] &mass range [$M_\odot$] &mass range [$M_\odot$]\\

\hline\hline

{\bf Orion nebula cluster, ONC}\\ 

Muench {\it et al.} \cite{MLL00}
&$-0.35$      &$+1.25$      &$+2.35$\\
{\it magenta small open circles with central dot} 
&$0.02-0.08$  &$0.08-0.80$  &$0.80-63.1$\\
{\it magenta large open circles with central dot} 

&$+0.00$      &$+1.00$      &$+2.00$\\
&$0.02-0.08$  &$ 0.08-0.40$ &$0.4-63.10$\\ 

Hillenbrand \& Carpenter \cite{HC00} (HC00)
&$+0.43$\\
{\it magenta large thick open circle}
&$0.02-0.15$\\
\hspace{5mm}{\it  with central dot}\\

Luhman \cite{L00}
&$+0.70$\\
{\it magenta small thick open circle}
&$0.035-0.56$\\
\hspace{5mm}{\it  with central dot}\\

\hline

{\bf Pleiades}\\
Moraux {\it et al.} \cite{MBS01} &$+0.51\pm0.15$\\
{\it green circles with central dot}
&$0.04-0.30$\\

Hambly {\it et al.} \cite{Hambetal99}, from \cite{Netal01}
&$+0.56$             &$+2.67$\\
{\it green circles with central dot}
&$0.065-0.60$        &$0.6-10.0$ \\

\hline

{\bf $\mathbf{\sigma}$ Ori}\\
Bejar {\it et al.} \cite{Betal01}
&$0.8\pm0.4$\\
{\it green solid circle}
&$0.013-0.20$\\

\hline

{\bf M35}\\
Navascues {\it et al.} \cite{Netal01}
&$-0.88\pm0.12$     &$0.81\pm0.02$   &$2.59\pm0.04$\\
{\it green solid circle}$^1$
&$0.08-0.2$         &$0.2-0.8$       &$0.8-6.0$\\

\hline

{\bf IC 348}\\
Najita {\it et al.} \cite{Netal00} for MLR from \cite{BCAH98}
&$+0.5$\\
{\it green solid circle}
&$0.015-0.22$\\

\hline

{\bf NGC~2264}\\
Park {\it et al.} \cite{Petal00}
&&&$+2.7$\\
{\it green solid circle}
&&&$2.0-6.3$\\

\hline

{\bf 5~LMC regions}\\
Parker {\it et al.} \cite{Petal01}
&&&$+2.3\pm0.2$\\
{\it blue solid triangle}
&&&$5-60$\\

\hline

{\bf NGC~1818} in LMC\\
Santiago {\it et al.} \cite{Santetal01}, outer region
&&$+2.5$\\
{\it blue solid triangle}
&&$0.9-3$\\

{\bf NGC~1805} in LMC\\
Santiago {\it et al.} \cite{Santetal01}, outer region
&&$+3.4$\\
{\it blue solid triangle}
&&$0.9-3$\\

\hline

\end{tabular}
}
\caption{\small{\it continued}}
\end{table}

\addtocounter{table}{-1}

\begin{table}
{\small
\begin{tabular}{lccc}

\hline

&$\alpha$   &$\alpha$   &$\alpha$\\
&mass range [$M_\odot$] &mass range [$M_\odot$] &mass range [$M_\odot$]\\

\hline\hline

{\bf 30~Dor}$^\star$ in LMC\\
Selman {\it et al.} \cite{Setal99}, $r>3.6$~pc
&&&$+2.37\pm0.08$\\
{\it cyan small open triangle}
&&&$3-120$\\

Selman {\it et al.} \cite{Setal99}, $1.1<r/{\rm pc}<4.5$
&&&$+2.17\pm0.05$\\
{\it cyan small open triangle}
&&&$2.8-120$\\

Sirianni {\it et al.} \cite{Setal00}
&&$+1.27\pm0.08$   &$+2.28\pm0.05$\\
{\it cyan large open triangle}$^2$
&&$1.35-2.1$       &$2.1-6.5$\\

\hline

{\bf Arches cluster}$^\star$\\
Figer {\it et al.} \cite{Fetal99}, all radii
&&&$+1.6\pm0.1$\\
{\it cyan large solid circle} 
&&&$6.3-125$\\

\hline

{\bf NGC~3603}$^\star$\\
Eisenhauer {\it et al.} \cite{Eetal98} 
&&$+1.73$    &$+2.7$\\
{\it cyan small solid circle}
&&$1-30$     &$15-70$\\

\hline

{\bf Globular clusters}\\
Piotto \& Zoccali \cite{PZ99}
&$+0.88\pm0.35$   &$+2.3$\\
{\it yellow open triangles} &$0.1-0.6$        &$0.6-0.8$\\

\hline

{\bf Galactic bulge}\\
Holtzman {\it et al.} \cite{Hetal98}
&$+0.9$         &$+2.2$\\
{\it magenta filled square}
&$0.3-0.7$      &$0.7-1.0$\\

Zoccali {\it et al.} \cite{Zetal00}
&$+1.43\pm0.13$ &$+2.0\pm0.23$\\
{\it magenta filled square}
&$0.15-0.5$     &$0.5-1.0$\\

\hline

\multicolumn{4}{l}{
{\bf Solar Neighborhood} ({\it magenta dotted lines})} \\
Reid {\it et al.} \cite{Retal99}
&$+1.5\pm0.5$\\
&$0.02-0.08$\\

Herbst {\it et al.} \cite{Hetal99}
&$\le +0.8$\\
&$0.02-0.08$\\

Chabrier \cite{Ch01a,Ch01b}
&$\le+1$      &$+1$        \quad / \quad $+2$\\
&$0.01-0.08$  &$0.10-0.35$ \quad / \quad $0.35-1.0$\\

\hline

\end{tabular}
}

\caption{\small{$\alpha(<\!lm\!>)$ data obtained after 1998. The data
are shown in Fig.~\ref{fig:apl} in addition to the previously
available data set compiled by Scalo \cite{Sc98}. Each $\alpha$ value
is obtained at $<\!lm\!>=(lm_2-lm_1)/2$, $lm\equiv {\rm log}_{10}m$,
by the respective authors by fitting a power-law MF over the
logarithmic mass range given by $m_1$ and $m_2$ listed above.  Some
authors do not quote uncertainties on their $\alpha$ values.  Notes:
$^\star$ are starburst clusters; $^1$ thin green open circle
emphasizes the low-mass M35 datum; $^2$ the mass range
$1.35<m/M_\odot<2.1$ may be incomplete and is emphasized by the cross
through the cyan large open triangle.  }}
\label{tab:apl}
\end{table}




\begin{figure}
\begin{center}
\rotatebox{0}{\resizebox{0.7 \textwidth}{!}{\includegraphics{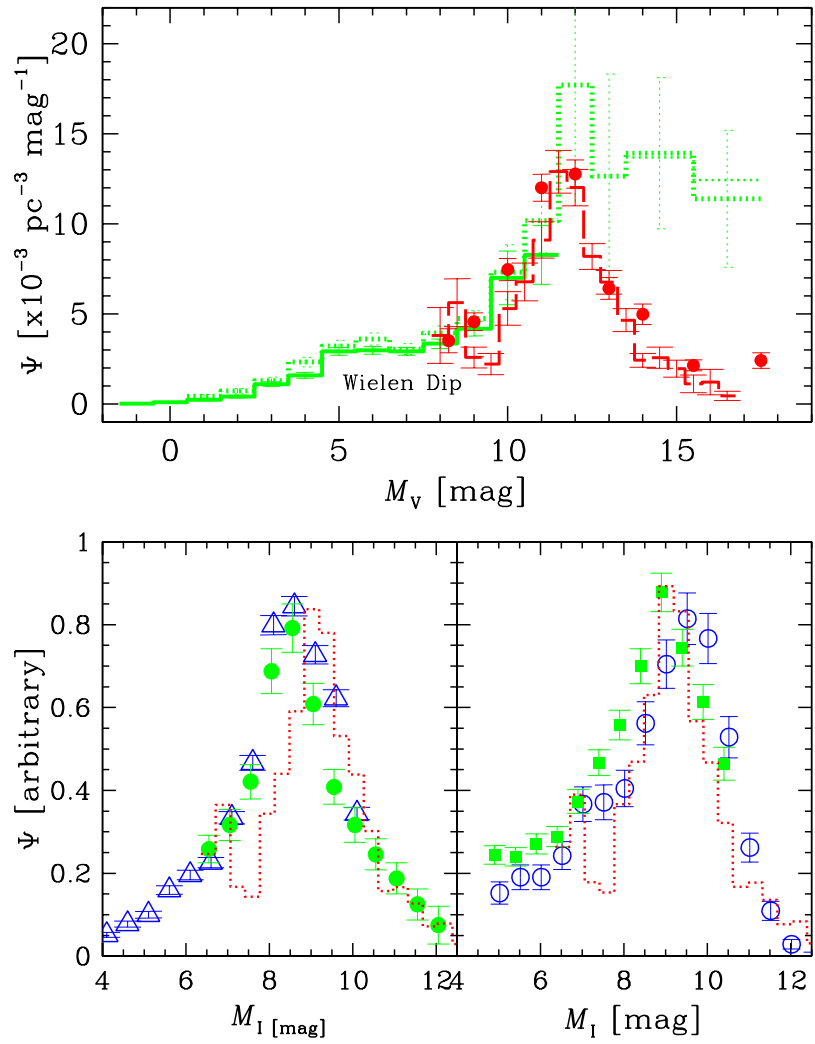}}}
\caption{\small{ Next page ....}}
\label{fig:lfs}
\end{center}
\end{figure}

\newpage

\begin{figure}
\begin{center}
\rotatebox{0}{\resizebox{0.75 \textwidth}{!}{\includegraphics{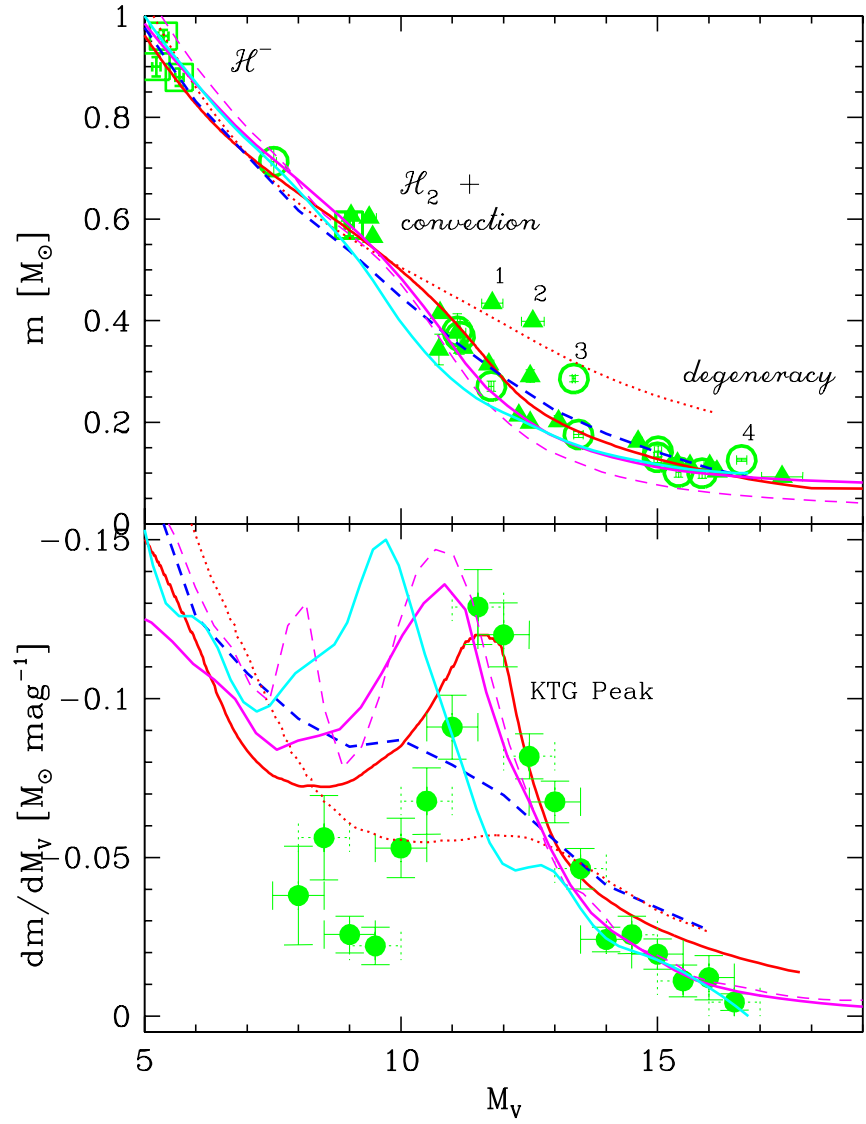}}}
\caption {\small{ Next page ...}}
\label{fig:mlr}
\end{center}
\end{figure}

\noindent
{\small{Fig.~\ref{fig:lfs}: Stellar luminosity functions (LFs, number
of stars per volume element and magnitude interval) for
solar-neighborhood \cite{neighb} stars as a function of absolute
magnitude in the V-band (upper panel) and four star clusters as a
function of absolute magnitude in the I-band (lower panel).  {\bf
Upper panel}: The photometric LF corrected for Malmquist bias
\cite{LFphot} and at the midplane of the Milky Way disk ($\Psi_{\rm
phot}$, red histogram) is compared with the nearby LF ($\Psi_{\rm
near}$, green histograms) constructed from the solar neighborhood
stellar sample \cite{neighb}. The average, ground-based
$\overline{\Psi}_{\rm phot}$ (dashed histogram, data pre-dating 1995
\cite{K95a}) is confirmed by Hubble-Space-Telescope (HST) star-count
data which pass through the entire Galactic disk and are thus not
prone to Malmquist bias (solid circles, \cite{Zheng01}). The
ground-based volume-limited trigonometric-parallax sample (dotted
histogram) systematically overestimates $\Psi_{\rm near}$ due to the
Lutz-Kelker bias \cite{LK}, thus lying above the improved estimate
provided by the Hipparcos-satellite data (solid histogram,
\cite{JW97,K01b}). The depression/plateau near $M_V=7$ is the {\it
Wielen dip}, named after Roland Wielen who's estimate of the LF in the
1970's for the first time unambiguously showed this feature.  The thin
dotted histogram at the faint end indicates the level of refinement
provided by recent stellar additions \cite{K01b} demonstrating that
even the immediate neighborhood within 5.2~pc of the Sun probably
remains incomplete at the faintest stellar luminosities. {\bf Lower
panel:} $I$-band LFs of stellar {\it systems} (single stars and
unresolved binaries) in four star clusters: the globular cluster (GC)
M15 \cite{deMP95a} (distance modulus \cite{magn} $\Delta
m=m-M=15.25$~mag, blue triangles), GC NGC~6397 \cite{PdeMR95} ($\Delta
m=12.2$, green solid circles), the young open cluster Pleiades
\cite{HJH91} ($\Delta m=5.48$, blue open circles), and the GC 47~Tuc
\cite{deMP95b} ($\Delta m=13.35$, green solid squares).  The dotted
histogram is $\overline{\Psi}_{\rm phot}(M_I)$ from the upper panel,
transformed to the $I$-band using the linear color--magnitude relation
$M_V=2.9+3.4\,(V-I)$ \cite{KTG93} and $\Psi_{\rm phot}(M_I) =
(dM_V/dM_I) \, \Psi_{\rm phot}(M_V)$.  The agreement in position and
amplitude of the maximum in the LFs for the five different populations
is impressive. This maximum results from a minimum in the derivative
of the mass--luminosity relation (Fig.~\ref{fig:mlr}). {\footnotesize
{\it Note}: The figure which appeared in Science issue of 4th January
2002 has a slightly erroneous lower panel. The version shown here is
corrected}.}}

\newpage

\noindent
{\small{Fig.~\ref{fig:mlr}: The mass--luminosity relation (MLR, upper
panel) is the mass of a star as a function of its absolute magnitude
in the V-band. The derivative of the MLR is plotted in the lower
panel. {\bf Upper panel:} The most recent observational data (solid
triangles and open circles, Delfosse et al., \cite{Detal00}; open
squares, Andersen, \cite{A91}) are compared with the empirical MLR of
Scalo (blue dashed line \cite{Sc86}) and the semi-empirical KTG93 MLR
(red solid curve \cite{KTG93}). The under-luminous data points~1--4
are metal-rich stars \cite{Detal00}.  The magenta solid line is a 5~Ga
old isochrone and the magenta dashed line is a 0.1~Ga isochrone for
solar metal abundances from Baraffe et al. \cite{BCAH98}. The cyan
solid line is a 5~Ga isochrone for metalicity $Z=0.02\,Z_\odot$ from
Siess et al. \cite{SDF00}. As the mass of a star is reduced, H$^-$
opacity becomes increasingly important through the short-lived capture
of electrons by H-atoms. This results in reduced stellar luminosities
for intermediate and low-mass stars. The $m(M_V)$ relation becomes
less steep in the broad interval $3<M_V<8$ leading to the Wielen dip
(Fig.~\ref{fig:lfs}).  The $m(M_V)$ relation steepens near $M_V=10$
because the formation of H$_2$ in the very outermost layers of
low-mass stars increases the mean molecular weight there causing the
onset of convection up to and above the photosphere. This leads to a
flattening of the temperature gradient and therefore to a larger
effective temperature, as opposed to an artificial case without H$_2$
but the same central temperature.  Brighter luminosities result. Full
convection establishes throughout the whole star for $m <
0.35\,M_\odot$.  The modern ML data beautifully confirm the steepening
in the interval $10<M_V<13$ predicted in 1990 \cite{KTG90}. The red
dotted MLR demonstrates the effect of suppressing the formation of the
H$_2$ molecule by lowering it's dissociation energy from 4.48~eV to
1~eV.  The $m(M_V)$ relation flattens again for $M_V>14$,
$m<0.2\,M_\odot$ as degeneracy in the stellar core becomes
increasingly important for smaller masses limiting further contraction
\cite{HN63,ChB97}. {\bf Lower panel:} The derivatives of the same
relations plotted in the upper panel are compared with
$\overline{\Psi}_{\rm phot}$ from Fig.~\ref{fig:lfs} scaled to fit
this figure.

\begin{figure}
\begin{center}
\rotatebox{0}{\resizebox{1.0 \textwidth}{!}{\includegraphics{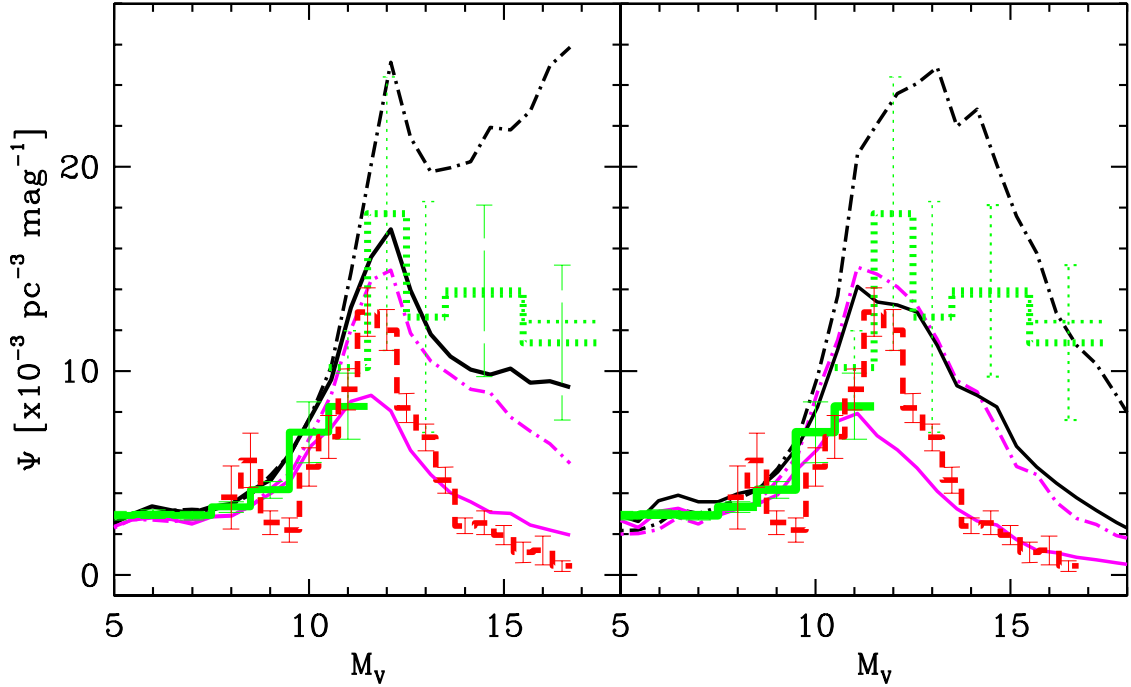}}}
\vskip -11cm
\caption
{\small{Model LFs (number of stars per unit volume and magnitude as a
function of the absolute magnitude in the V-band) are constructed
using the semi-empirical KTG93 MLR \cite{KTG93} (left panel) and the
most advanced theoretical MLR computed by Baraffe et al. for a 5~Ga
population of solar composition \cite{BCAH98} (right panel). The MLRs
are plotted in Fig.~\ref{fig:mlr}.  The models are compared with the
observed solar-neighborhood LFs shown in Fig.~\ref{fig:lfs}.  For a
given IMF, the upper (black) curves are single-star LFs. The lower
curves show the unresolved system LFs in which the luminosities of
stellar companions are added for a population of 8000 single stars,
8000 binaries, 3000 triples and 1000 quadruples (40:40:15:5~\%,
respectively). Companions with masses $0.08\le m/M_\odot\le 1$ are
combined randomly from the IMF. The models assume perfect photometry,
no distance errors and no metalicity or age spread. The model system
LFs thus reflect the empirical photometric LF corrected for Malmquist
bias, $\Psi_{\rm phot}$, whereas the observed $\Psi_{\rm near}$ is
broadened mostly due to the metalicity and partially an age spread
which is not modeled.  The models are scaled to fit the LFs at
$M_V\approx7$ with equal scaling for the single-star and system LFs
for a given IMF.  In the left panel the IMF is is a two-component
power-law with Salpeter exponent $\alpha_2=2.3$ for $0.5-1.0\,M_\odot$
but for $0.08-0.5\,M_\odot$, $\alpha_1=1.6$ for the dot-dashed model
and $\alpha_1=1.0$ for the solid model.  In the right panel it is a
one-component power-law, $\xi(m)\propto m^{-\alpha}$, over the whole
mass range ($0.08-1\,M_\odot$) with $\alpha=1.8$ (dot-dashed model)
and $\alpha=1.2$ (solid model).  The models are selected to roughly
give similar overall deviations about the data and are not intended to
be best-fit solutions. Note that the change in shape of the LF,
$d^2\Psi/dM_V^2$, is an interesting observable containing information
about the MLR and the underlying IMF.}}
\label{fig:lfmods}
\end{center}
\end{figure}

\begin{figure}
\begin{center}
\rotatebox{0}{\resizebox{1.0 \textwidth}{!}{\includegraphics{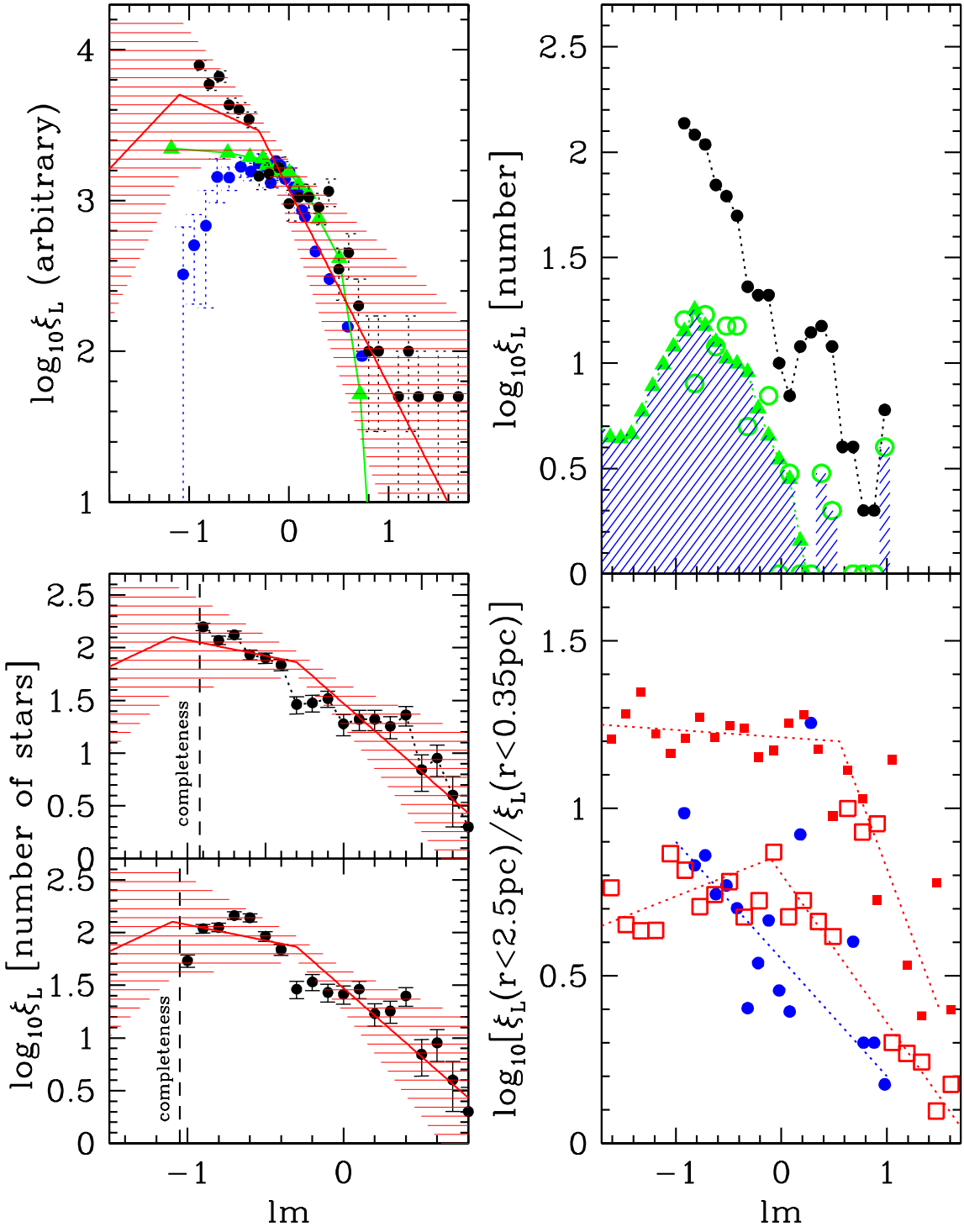}}}
\vskip -10mm
\caption
{\small{Next page...
}}
\label{fig:mfn}
\end{center}
\end{figure}


\newpage

\vskip 0mm
\noindent
{\small{Fig.~\ref{fig:mfn}: {\bf Upper left panel:} The measured
stellar mass functions, $\xi_{\rm L}$, as a function of logarithmic
stellar mass ($lm\equiv {\rm log}_{10}(m/M_\odot)$) in the Orion
nebula cluster (ONC, solid black circles, \cite{HC00}), the Pleiades
(green triangles, \cite{Hambetal99}) and the cluster M35 (blue solid
circles, \cite{Netal01}). The decrease of the M35 MF below
$m\approx0.5\,M_\odot$ remains present despite using different MLRs.
None of these MFs are corrected for unresolved binary systems.  The
average Galactic-field single-star IMF is shown as the solid red line
with the associated uncertainty range (eq.~\ref{eq:imf} in
Table~\ref{tab:imfs}). The ONC data are from the Hillenbrand optical
survey within $r=2.5$~pc of the center of the cluster. The cluster is
$\tau<2$~Ma old and has a metalicity [Fe/H]$\;=-0.02$. For the
Pleiades, $r=6.7$~pc, $\tau\approx100$~Ma and [Fe/H]$\;=+0.01$. For
M35 $r=4.1$~pc, $\tau_{\rm cl}\approx160$~Ma and
[Fe/H]$\;=-0.21$. {\bf Lower left panel:} The shape of the ONC MF
differs for very low-mass stars above the completeness limit of the
survey if different pre-main sequence evolution tracks, and thus
essentially different theoretical MLRs by D'Antona \& Mazzitelli (DM)
are employed. For more details see \cite{HC00}.  The lower part shows
the ONC MF if ``DM94'' pre-main sequence models are used, whereas the
upper part shows the MF if ``DM97/98'' models are used. The average
IMF is as in the upper left panel. {\bf Upper right panel:} Mass
segregation is very pronounced in the ONC. This is evident by
comparing the MF for all stars within two different radial regions
centered on the cluster center.  The solid black circles are for all
stars within $r=2.5$~pc and the open green circles are for all stars
within $r=0.35$~pc, from the Hillenbrand ONC survey \cite{H97}. The
solid green triangles are for $r=0.35$~pc, from \cite{HC00}.  {\bf
Lower right panel:} The ratio of the MFs in the different circular
survey regions of the upper right panel shows the pronounced mass
segregation in the ONC.  The IMF ratio, $\xi_{\rm L}(r<2.5\,{\rm
pc})/\xi_{\rm L}(r<0.35\,{\rm pc})$, is plotted as blue solid circles.
It increases with decreasing mass. This comes about because the number
of low-mass stars is depleted in the inner ONC region.
Stellar-dynamical models of the ONC can be used to study if the
observed mass segregation (blue solid dots) can be arrived at by
dynamical mass segregation. If not, then we have definite proof that
the mass segregation is primordial and thus that the IMF varies at
least on small scales ($<1$~pc).  The model snapshots shown are from
model~B in \cite{KAH} and assume the average IMF.  The masses of
single stars and binary systems are counted to construct $\xi_{\rm
L}$.  Initially the ratio is constant with stellar mass because the
model starts with no mass segregation.  The red solid squares are a
snapshot at 0.9~Ma, whereas the red open squares are for 2.0~Ma.  The
dotted lines are eye-ball fits to the data. The data demonstrate that
mass segregation develops rapidly and that by about 2~Ma the observed
effect is obtained. This casts doubt on the primordial origin of the
observed mass segregation.  }}

\voffset -3cm
\begin{figure}
\begin{center}
\rotatebox{0}{\resizebox{1.0 \textwidth}{!}{\includegraphics{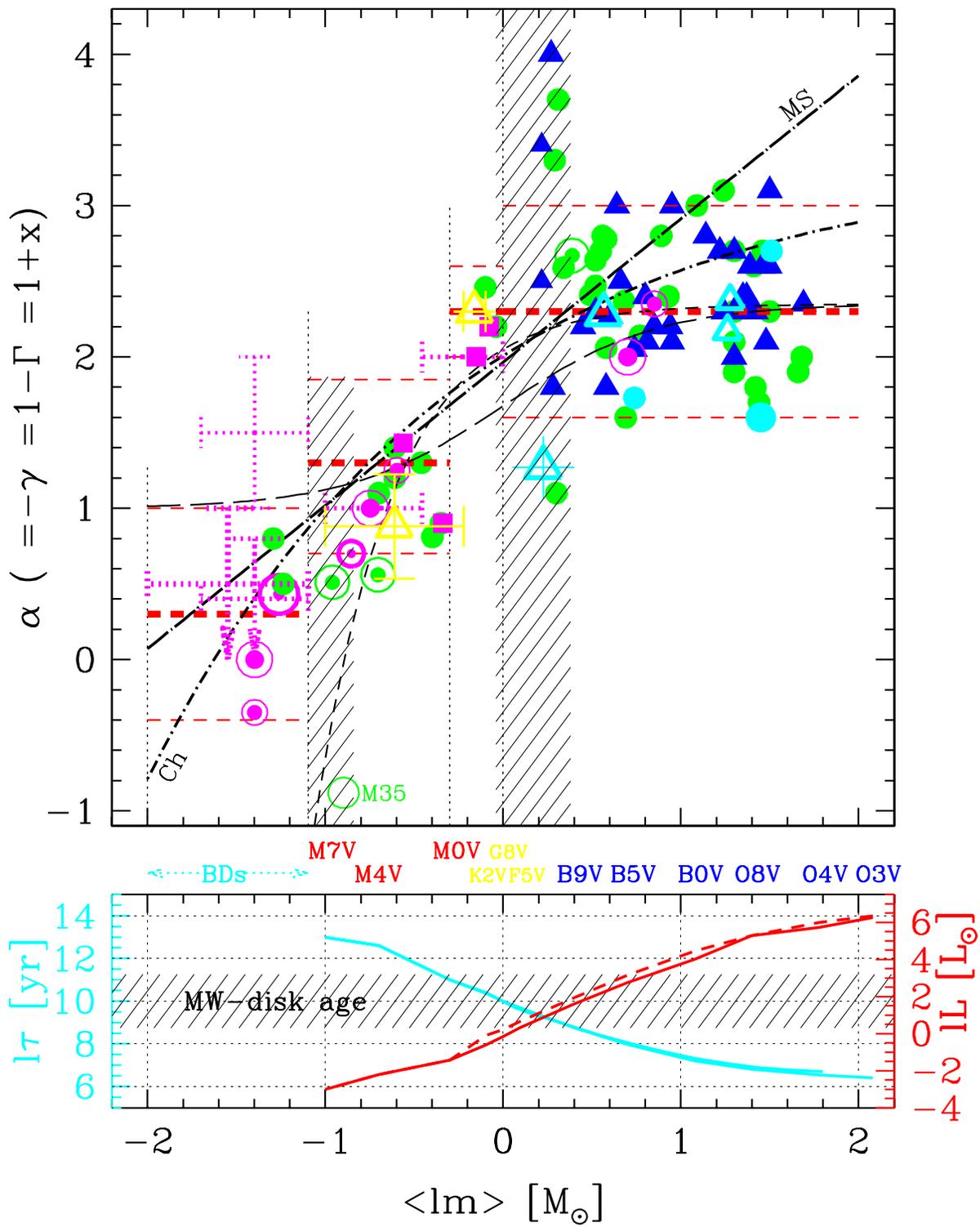}}}
\vskip -8mm
\caption[]{\small{ Next page...}}
\label{fig:apl}
\end{center}
\end{figure}

\vfill
\newpage

\noindent
{\small{Fig.~\ref{fig:apl}: {\bf Upper panel:} The alpha plot compiles
measurements of the power-law index, $\alpha$, as a function of the
logarithmic stellar mass and so measures the shape of a MF.
[Notation: $lm\equiv {\rm log}_{10}(m/M_\odot)$, $l\tau\equiv {\rm
log}_{10}(\tau/{\rm yr})$, $lL\equiv {\rm log}_{10}(L/L_\odot)$].  The
shape of the MF is mapped in the upper panel by plotting measurements
of $\alpha$ at $<\!lm\!>=(lm_2-lm_1)/2$ obtained by fitting
power-laws, $\xi(m)\propto m^{-\alpha}$, to logarithmic mass ranges
$lm_1$ to $lm_2$ (not indicated here for clarity).  Many of the green
circles and blue triangles are pre-1998~data compiled by Scalo
\cite{Sc98,K01a} for MW (green filled circles) and
Large-Magellanic-Cloud clusters and OB associations (blue solid
triangles). Newer data are also plotted using the same symbols, but
some are emphasized using different symbols and colors, such as by
yellow triangles for globular cluster MFs (Table~\ref{tab:apl}).
Unresolved multiple systems are not corrected for in all these data
including the MW-bulge data.  The average solar-neighborhood IMF
(eq.~\ref{eq:imf} in Table~\ref{tab:imfs}) are the red thick
short-dashed lines together with the associated uncertainty ranges.
Other binary-star-corrected solar-neighborhood-IMF measurements are
indicated as magenta dotted error-bars (Table~\ref{tab:apl}).  The
quasi-diagonal black lines are analytical forms summarized in
Table~\ref{tab:imfs}. The vertical dotted lines delineate the four
mass ranges (eq.~\ref{eq:imf} in Table~\ref{tab:imfs}), and the shaded
areas highlight those stellar mass regions where the derivation of the
IMF is additionally complicated especially for Galactic field stars:
for $0.08<m/M_\odot< 0.15$ long pre-main sequence contraction times
\cite{ChB00} make the conversion from an empirical LF to an IMF
(eq.~\ref{eq:mf_lf}) dependent on the precise knowledge of stellar
ages and the SFH.  For $0.8< m/M_\odot<2.5$ uncertain main-sequence
evolution, Galactic-disk age and the SFH of the MW disk do not allow
accurate IMF determinations \cite{Binney00}.  {\bf Lower panel:} The
bolometric MLR, $lL(lm)$, and stellar main-sequence life-time,
$l\tau$, are plotted as a function of logarithmic stellar mass.  The
uncertainty in the age of the Milky-Way disk is shown as the shaded
region. Stellar spectral types are written between the panels. }}


\begin{figure}
\begin{center}
\rotatebox{90}{\resizebox{1.0 \textwidth}{!}{\includegraphics{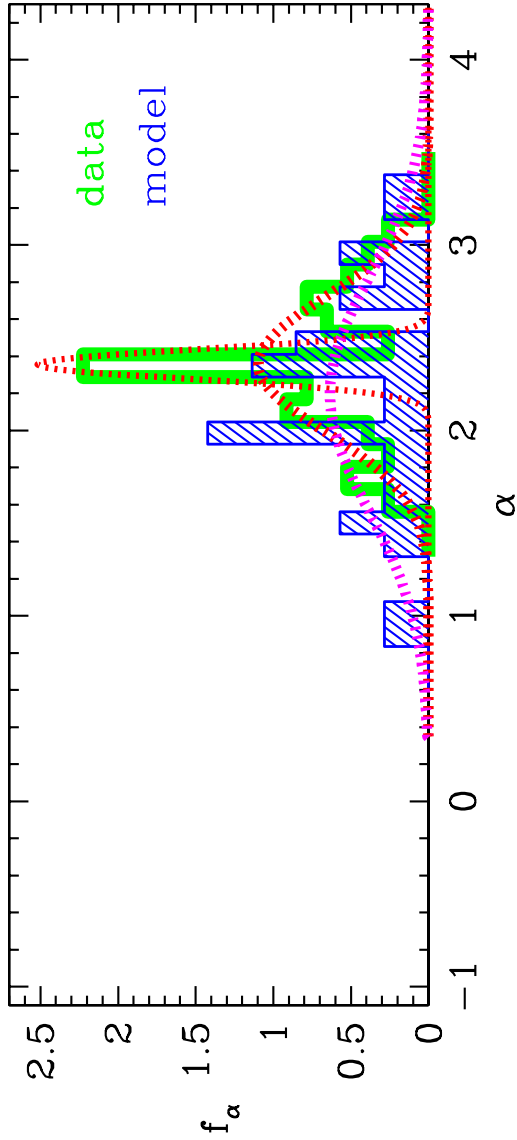}}}
\vskip 0mm
\caption
{\small{The histogram of MF power-law indices ($\alpha$) for
massive stars ($lm>0.40$).  If the $\alpha$ measurements are not
distributed like a Gaussian function then this may imply that some of
the data are different from the mean because of true IMF variations.
The green histogram shows the observational data from
Fig.~\ref{fig:apl}.  The blue shaded histogram shows theoretical
values from an ensemble of 12 star clusters containing initially 800
to $10^4$~stars that are snapshots at 3 and 70~Ma \cite{K01a}. Stellar
companions in binaries are merged to give the system MFs, which are
used to measure $\alpha$.  The assumed IMF is eq~\ref{eq:imf} in
Table~\ref{tab:imfs}.  The dotted curves are Gaussians with mean
$\alpha$ and standard deviation, $\sigma_\alpha$, obtained from the
histograms. The theoretical data give $<\!\!\alpha\!\!> = 2.20,
\sigma_\alpha=0.63$ (magenta dotted curve), and thus arrive at the
input Salpeter value.  The empirical data from Fig.~\ref{fig:apl} give
$<\!\!\alpha\!\!> = 2.36, \sigma_\alpha=0.36$ which is the Salpeter
value.  Fixing $\alpha_{\rm f}=<\!\!\alpha\!\!>$ and using only
$\!\mid \alpha \!\mid \le 2\,\sigma_\alpha$ for the observational data
gives the narrow thin red dotted Gaussian distribution which describes
the Salpeter peak ($\alpha_{\rm f}=2.36, \sigma_{\alpha,f}=0.08$).
}}
\label{fig:ahist}
\end{center}
\end{figure}

\end{document}